\documentclass[universe,review,accept,oneauthor,pdftex]{Definitions/mdpi} 

\firstpage{1} 
\pubvolume{8}
\issuenum{2}
\articlenumber{90}
\pubyear{2022}
\copyrightyear{2022}
 \externaleditor{Academic Editor:{Francesca Calore}} 
\datereceived{24 December 2021 } 
\dateaccepted{26 January 2022 } 
\datepublished{29 January 2022 } 
\hreflink{https://doi.org/10.3390/\linebreak universe8020090} 

\Title{\textls[-5] {Statistical Tools for Imaging Atmospheric Cherenkov~Telescopes}}
\TitleCitation{Statistical Tools for Imaging Atmospheric Cherenkov Telescopes}

\Author{Giacomo D'Amico \orcidA{}}

 \AuthorNames{Giacomo D'Amico}

 \AuthorCitation{D'Amico, G.}

\address[1]{%
 Department for Physics and Technology, University of Bergen,  NO-5020 Bergen, Norway; giacomo.damico@uib.no
}

\abstract{The development of Imaging Atmospheric
Cherenkov Telescopes (IACTs) unveiled the sky in the teraelectronvolt regime, initiating the so-called ``TeV revolution'', at the beginning of the new millennium. This revolution was also facilitated by  the implementation and adaptation of statistical tools for analyzing the shower images collected by these telescopes and inferring the properties of the astrophysical sources that produce such events. Image reconstruction techniques, background discrimination, and signal-detection analyses are just a few of the pioneering studies applied in recent decades in the analysis of IACTs data.   This (succinct) review has the intent of summarizing the most common statistical tools that are used for analyzing data collected with IACTs, focusing on  their application in the full analysis chain,  including references to existing literature for a deeper~examination. 
 }

\keyword{statistical analysis; gamma rays; IACTs; likelihood; bayes} 

\begin{document}


\section{Introduction}

Any scientific experiment would be incomplete if only the collected data were reported. A~statistical analysis is needed in order to interpret the data and to draw conclusions from the experiment. This is  the case for experiments that are imaging the Cherenkov light emitted by a cascade of secondary particles produced by the interaction of gamma rays and cosmic rays within the Earth's atmosphere. They are called Imaging Atmospheric Cherenkov telescopes  and popular examples are MAGIC~\cite{aleksic2016major}, HESS~\cite{hinton2004status}, VERITAS~\cite{holder2006first}, and CTA~\cite{actis2011design}, which is currently under construction.   
By ``statistic''  we mean any function $\mathcal{S}$\endnote{Hereafter throughout the paper the symbol $\mathcal{S}$ is used for the statistic,  while the generic symbol $ p()  $ is used to indicate all probability density functions (PDFs) and probability mass functions (PMFs) (the former applies to continuous variables and the latter to discrete variables).} computed from the observed data assuming the truth of a model. Very well-known examples of such functions are the mean, the~variance and the $\chi^2$. As the observed data consists of random variables, the~statistic itself is a random variable whose distribution can be derived either from theoretical considerations or empirically using Monte Carlo (MC) simulations. A~statistical analysis is therefore performed by comparing the observed value of the statistic with the frequency distribution of the values of the statistic from hypothetical infinite repetition of the same experiment assuming a given model of interest. This approach is usually called the ``classical'' or ``frequentist'' approach.  This comparison (referred to as the test statistic) between the observed statistic and its long-run distribution allows the analyzer to draw a conclusion from the observed data with a procedure that is right\endnote{By convention $\alpha$ is the probability of making a type I error, i.e.,~rejecting a hypothesis that is true. It is also refereed as the statistical significance or \emph{p}-value.} $(1-\alpha) \cdot 100 \%$ of the time. The~value $(1-\alpha) \cdot 100 \%$  is referred to as the confidence level (CL). 
It is important to underline that it is the procedure, not the conclusion, which is correct $(1-\alpha) \cdot 100 \%$ of the time. 
To better clarify this point we can consider the following claim: ``a flux of $10^{-13} \cdot \text{cm}^{-2} \text{\; s}^{-1}$   from the observation of a gamma ray burst is  excluded at $95\% $ CL''. 
Claiming that such a value of the flux is excluded is obviously always wrong in infinite experiments in which the true flux of the observed gamma ray burst is $10^{-13} \cdot \text{cm}^{-2} \text{\; s}^{-1}$. However, in these infinite experiments, the procedure would lead the analyzer to this wrong conclusion only $5\%$ of the time\endnote{Here we are assuming that the analyzer would make this conclusion every time that the observed statistic falls above the 95th-percentile of the statistic distribution.}. The~procedure, i.e.,~the test statistic and the value $\alpha$ for its significance, is usually dictated by many factors, such as the assumptions about the underlying model, the~way the data have been collected and, sometimes, also by the biased conclusions\endnote{The misuse and misinterpretation of statistical tests in the scientific community led the American Statistical Association (ASA)  to release in 2016 a statement~\cite{doi:10.1080/00031305.2016.1154108} on the correct use of statistical significance and \emph{p}-values.} one is willing to derive from the experiment. A~common principle~\cite{neyman1933ix} is to choose the test statistic with the maximum power, where the power of the statistic is the probability of rejecting a hypothesis that is false. According to the Neyman–Pearson lemma the most powerful statistic is the likelihood ratio, usually defined in the literature, for~reasons that will be clear soon, as~follows
\begin{equation}
    -2 \log  \frac{ \mathcal{L}(  \theta | D_{obs} ) }{\mathcal{L}( \hat{\theta} | D_{obs} ) },
    \label{Eq:LKratio}
\end{equation}
which by definition can only take values bigger or equal than zero, since by $\hat{\theta}$ we have defined  the values of the model parameters that maximize the likelihood $\mathcal{L}$.
The likelihood  is a function of the model parameters $\theta$ defined as the probability of obtaining the observed data $D_{obs}$ assuming $\theta$ to be true:
\begin{equation}
    \mathcal{L}(  \theta | D_{obs} ) = p( D_{obs} | \theta).
    \label{Eq:likelihood}
\end{equation}

 Searching for the parameter values $\hat{\theta}$ that maximize the likelihood in Equation~(\ref{Eq:likelihood}) is also referred to as fitting the model to the data.  If~nuisance parameters\endnote{By nuisance parameters we mean parameters that are not of interest but must be accounted for.} $\pi$ are present in the model, $\pi$ is maximized to the value $\hat{\hat{\pi}}$ in the numerator of Equation~(\ref{Eq:LKratio}) letting $\theta$ be free, resulting in the so-called likelihood profile
\begin{equation}
        \mathcal{L}(  \theta | D_{obs} ) = \mathcal{L}(  \theta , \hat{\hat{\pi}}(\theta) | D_{obs} ).
        \label{Eq:Lkl_profile}
\end{equation}

Taking the log value of the likelihood ratio as done in Equation~(\ref{Eq:LKratio}) allows making use of Wilk's theorem~\cite{Wilks}, which states that under certain circumstances this random variable has a $\chi^2$ distribution with degrees of freedom given by the number of the free parameters. This property makes the likelihood ratio a very appealing statistic, whose usage has a very broad application, and~it will indeed appear many times in this manuscript.  Yet, it must be used   cautiously: the interested reader may refer to Ref.~\cite{protassov2002statistics} for a critical review of the usage and abuse of the likelihood~ratio. 

The frequentist theory  described so far may be considered  unsatisfactory by some~\cite{loredo1990laplace}  with its dependence on long-run distributions from infinite experiments and its arbitrariness in the choice of the statistic. An~alternative approach is provided by the ``Bayesian'' or ``probabilistic'' approach, in~which probabilistic statements about hypotheses and model parameters are made through the Bayes theorem
\begin{equation}
    p( \theta | D_{obs} ) = \frac{p( D_{obs} | \theta ) p(\theta) }{ p(D_{obs})}.
    \label{Eq:Bayes}
\end{equation}

The prior probability $p(\theta)$ captures the available knowledge about the parameters, or, more generally, about the hypotheses under study. The~so-called evidence $p(D_{obs})$ can be seen as a normalization factor. It follows from probability theory that
\begin{equation}
    p(D_{obs}) = \sum_{\theta} p( D_{obs} | \theta ) p(\theta).
\end{equation}

In this case nuisance parameters are treated via the marginalization
\begin{equation}
   p( \theta | D_{obs} ) = \sum_{\pi} p( \theta, \pi | D_{obs} ) ,
\end{equation}
or, in~other words, instead of profiling the likelihood by fixing $\pi$ to $\hat{\hat{\pi}}(\theta)$,  one marginalizes the likelihood  by integrating out $\pi$. Another way of looking at Equation~(\ref{Eq:Bayes}) is to consider the odds of a hypothesis defined as the ratio of its probability of being true and not being~true
\begin{equation}
    o( \mathcal{H} ) = \frac{p(\mathcal{H})}{1- p(\mathcal{H})} \equiv  \frac{p(\mathcal{H})}{ p(\bar{\mathcal{H}})},
\end{equation}
where $\bar{\mathcal{H}} $ and $\mathcal{H}$ are mutually exclusive and collectively exhaustive hypotheses. Using the odds formalism the Bayes theorem takes the following form
\begin{equation}
    o( \mathcal{H} | D_{obs} ) = \text{BF} \cdot o( \mathcal{H}  ) , \quad \text{with}  \quad \text{BF} = \frac{  p( D_{obs} | \mathcal{H} )}{ p(D_{obs} | \bar{\mathcal{H}})},
\end{equation}
where BF stands for the Bayes Factor, i.e.,~the likelihood ratio of two competing hypotheses. After~the measurement the Bayesian approach leads us to update   the odds we assign to a given hypothesis by multiplying it with the BF. Unlike the frequentist approach where the goal is to provide a statement about the long-run performance of a test statistic, in~Bayes theory we are not interested in hypothetical infinite experiments but in calculating the probability of hypotheses from the observed data and from our prior knowledge of~them.

Going deep into the details of the statistical analysis and the difference between the frequentist and Bayesian theory goes beyond the scope of this paper, the~interested readers may refer to Refs.~\cite{lyons2008open,van2021bayesian} and references therein. Yet, this brief introduction of the basic principles and definitions used for performing an inference analysis in both the frequentist and Bayesian approach is  necessary for reviewing  the statistical tools used in IACTs analysis in the  next~sections.

\begin{figure}[H]
    
    \includegraphics[width=0.98\linewidth]{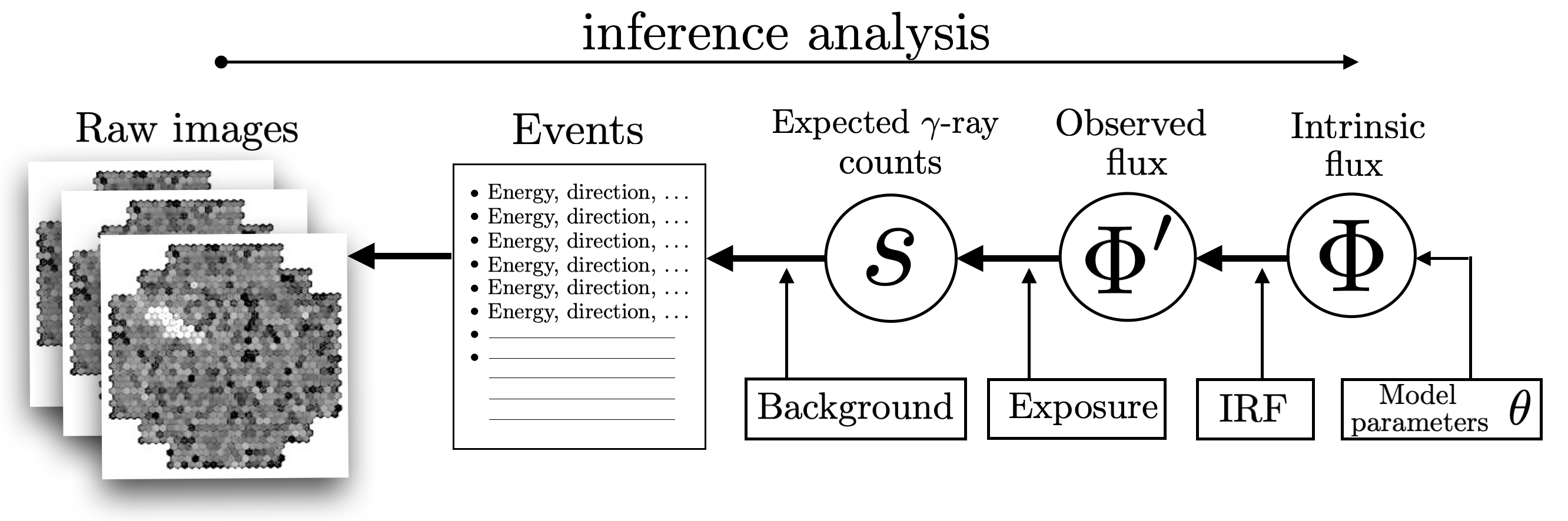}
    \caption{Schematic workflow of the inference analysis performed in order to estimate  the intrinsic flux $\Phi$ of gamma rays (and the values of its parameters $\theta$) from the recorded images. The~acronym IRF stands for Instrument Response Function (see Section~\ref{Sec:Flux}).  The~bold arrows (going from the right to the left)  show the relation of cause and effect. The~aim of the inference analysis (shown as a thin arrow going from the left to the right) is to invert such~relation. }
    \label{fig:Scheme}
\end{figure}

The typical workflow of an  IACTs analysis is schematically shown in Figure~\ref{fig:Scheme}, where, starting from the shower images, the~variable $s$ (the expected number of signal events) is derived, which is then used for inferring the flux $\Phi$ of gamma rays  by taking into account the instrument response function (IRF) of the~telescopes. 
 
In the remaining part of this section, the structure of the paper is outlined. First we discuss in Section~\ref{Sec:EventRecon} the most common techniques implemented for performing the event reconstruction from the shower images  detected with IACTs. The~goal of these  techniques' yields is to obtain a list with the estimated energy, direction, and discriminating variables for each of the candidate gamma ray~events. 

Using this event list, it is then shown in Section~\ref{Sec:detection} how to estimate the strength of the signal $s$ and how confidently we can claim that a gamma ray source is producing part of the recorded~events. 

The final result of the statistical analysis is the differential gamma ray flux $\Phi $, which corresponds to  the number $N_{\gamma} $ of expected photons per unit energy ($E$), time ($t$), and area~($A$):
\begin{equation}
    \Phi(E,t,\hat{\mathbf{n}}) = \frac{dN_{\gamma} (E,t,\hat{\mathbf{n}})}{dE d A dt},
    \label{Eq:InFlux}
\end{equation}
where $ \hat{\mathbf{n}}$ is the photon direction. We denote with $\Phi'$ the observed flux, i.e.,~the flux of events actually observed by the telescopes when the IRF of the telescope is included. The~expected counts $s$ is connected to $\Phi'$ by taking into account the exposure of the observation, i.e.,~by integrating $\Phi'$ over the temporal, energetic and spatial range  in which the events have been collected. This is the topic of Section~\ref{Sec:Flux}, in~which the way the source flux and its model parameters are obtained is also~discussed.

\section{Event Reconstruction~Techniques}
\label{Sec:EventRecon}

\begin{figure}[H]
    
    \includegraphics[width=0.9\linewidth]{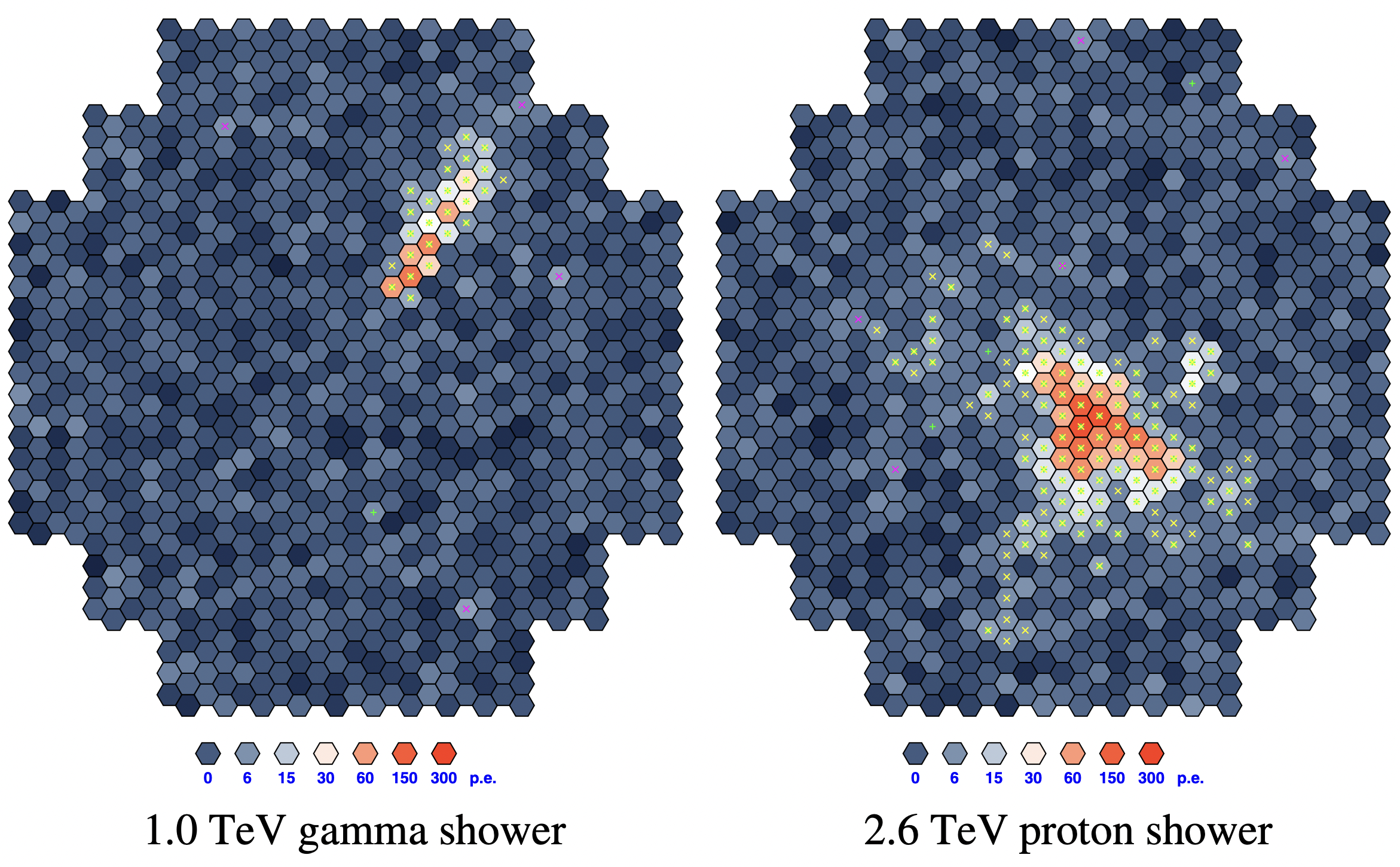}
    \caption{\textls[-5]{Difference between the images of gamma-induced (\textbf{left}) and hadron-induced (\textbf{right}) showers in the camera of a IACT. Reprinted with permission from Ref.~\cite{cite-key}.} Copyright 2009 Völk et al.}
    \label{fig:CameraImage}
\end{figure}

The first statistical analysis one has to face in IACTs is the reduction of the recorded images in the camera of the telescopes  to a few parameters of interest. The~Cherenkov light from the shower of secondary particles is reflected by mirrors and focused on a camera with photomultipliers composing the pixels of the shower image (see Figure~\ref{fig:CameraImage}). The~event reconstruction consists therefore in extracting from the photo-electron (PhE) counts and arrival time of each  pixel the following~variables:
\begin{itemize}
    \item the energy of the primary gamma ray that initiated the shower,
    \item its arrival direction,
    \item and one or more discriminating variables.
\end{itemize}

 The role of  these discriminating variables is to provide information on how likely one event can be associated with a gamma ray or to the background composed mainly by hadronic cosmic rays. The~background estimation and the signal extraction are discussed in Section~\ref{Sec:detection}, while for the remaining part of this section the most commonly used event reconstruction tools are~reviewed.

\unskip 

\subsection{Hillas~Method}
\label{Sec:Hillas}
The most common  event reconstruction technique is based on the moments (up to the second order) of the pixel amplitudes in the camera, referred to as Hillas parameters~\cite{1985ICRC....3..445H}. This technique can be thought of as fitting an ellipse to the pixels: a likelihood function that depends on the Hillas parameters is maximized under the assumption that the Cherenkov light from a shower initiated by a gamma ray would produce an elliptical shape in the camera.  The~set of parameters includes variables such as the total PhE counts in all pixels, the~PhE-weighted barycenter of the pixel positions, and the time gradient of the pixel arrival times. If~more than one telescope is involved, then these single-image parameters can be combined in order to obtain stereoscopic parameters giving a 3-dimensional reconstruction of the event~\cite{zanin2013mars}. The~calculation of these parameters is easily affected by image noise and night-sky background, which requires a cleaning procedure in order to remove the pixels that do not contain the shower image. Moreover, the~dim and small shower images below 100 GeV can result in parameters values affected by large fluctuations  and systematic uncertainties, which is the reason why the instrument response function of IACTs deteriorates at lower energies. Techniques based on Hillas parameters have been implemented since the 1980s and are used in a variety of experiments such as MAGIC~\cite{zanin2013mars} and HESS~\cite{aharonian2006observations}, demonstrating the robustness and reliability of the~method. 

After the parametrization of the event is completed, the~gamma ray energy is estimated from the shower impact parameter and from the photon density is measured
with each telescope. This is done by constructing look-up tables for different observational conditions, filled with MC information about the true energy of the gamma ray  as a function of the simulated image amplitude and  impact parameter. The~arrival direction is obtained  from the crossing point of the main ellipse orientations in the individual cameras. A~weighted combination of some Hillas parameters can be  used as a discriminating variable~\cite{aharonian2006observations}. More refined techniques have been developed, aimed at improving the inference analysis on the gamma ray properties starting from the Hillas parameters (see  Section~\ref{Sec:Multivar}).

\subsection{Semi-Analytical~Method}
\label{Sec:2D-model}
Despite the robustness and stability under different conditions of the Hillas method, additional reconstruction procedures have been explored in order to exploit more information from the recorded image.  The~so-called semi-analytical method consists of fitting  to the shower images  a model of the Cherenkov light produced by a gamma ray shower as seen by the camera. A~first implementation of this method can be found in Ref.~\cite{le1998new} from the CAT collaboration in the late 1990s. In~this pioneering implementation, the 2D-models are stored in a look-up table and compared to the observed image via a $\chi^2$-function of the gamma ray energy, the~impact parameter and the source position in the focal plane. This function is defined as the sum of the squared differences over all pixels between the expected PhE content and the actual observed one. This sum is weighted according to the Phe count quadratic error. A~$\chi^2$ minimization is performed to obtain the best fit parameter of the gamma ray, while the resulting $\chi^2$ is then used as a discriminating variable. This method has been re-implemented and subsequently improved by the HESS collaboration~\cite{de2009high}, where the $\chi^2$ minimization has been substituted by the minimization of a log-likelihood defined as
\begin{equation}
    \ln \mathcal{L} = \sum_{\text{pixels } i} \ln \mathcal{L}_i = -2 \sum_{\text{pixels } i}  \ln p( n_i | \theta ).
\end{equation}

 The variable $n_i$ is the observed PhE count in the pixel $i$, while $\theta$ are the shower model parameters.
This method is referred to as semi-analytical because the template library of shower images is produced with MC simulations which are usually carried out with dedicated software such as KASKADE~\cite{KERTZMAN1994629}  and CORSIKA~\cite{heck1998corsika}. Compared with the Hillas method, a~more precise estimation of the energy and direction of the primary gamma ray is provided by this reconstruction technique, especially at low~energies.

\subsection{3D-Gaussian~Model}
\label{Sec:3D-model}
An additional approach is given in Refs.~\cite{lemoine2006selection, naumann2009upgrading}, where the single-pixel PhE counts are fitted to an analytical gaussian air shower model. This method, referred to as a 3D model or 3D Gaussian model, assumes an isotropic angular distribution  of the shower, and its rotational symmetry with respect to its incident direction is used to select gamma ray events. As~usual a likelihood function is maximized with respect to the shower parameters. This maximization process is rather fast thanks to the simpler assumption of the 3D-Gaussian model. More recently~\cite{becherini2011new} the 3D-model was combined with a multivariate analysis that makes use of the so-called \textit{Boosted Decision Tree} (see Section~\ref{Sec:Multivar}) and adapted to the detection necessity of IACTs, particularly for the discovery of new faint~sources.

\subsection{MC Template-Based~Analysis}
\label{Sec:MC}
The previously  mentioned methods in Sections~\ref{Sec:2D-model} and \ref{Sec:3D-model} strongly rely on a model fit that becomes more difficult to describe as we reach higher energies. The~more energetic the gamma ray, the~more particles are produced, and~a large fraction of the latter is capable of reaching the ground. This causes strong fluctuations in the fit model above $\sim$10 TeV. Moreover, in~these approaches, the quality of the model fit is inevitably worsened by instrumental effects and atmospheric conditions which require approximations in order to be taken into account. To~overcome these issues and improve the accuracy of the analysis, the~authors of Ref.~\cite{parsons2014monte} proposed an Image Pixel-wise fit for Atmospheric Cherenkov Telescopes (ImPACT). In~this approach, the template shower images are produced using more detailed and time consuming MC simulations. The~simulation chain consists  in simulating the air shower with CORSIKA~\cite{heck1998corsika} which is then combined with \textit{sim\_telarray}\endnote{The \textit{sim\_telarray} code is a program that given as input a complete set of photon bunches simulates the camera response of the telescope.}~\cite{bernlohr2008simulation}~for reproducing the instrumental effects of the telescopes. The~sensitivity is improved by a factor of 2 when the ImPACT reconstruction method is implemented, relative to the Hillas-based method (see Section~\ref{Sec:Hillas}). Compared with the 2D-model, some improvements were shown at higher energies.  A~similar implementation for the VERITAS telescopes can be found in Ref.~\cite{vincent2015monte}. The~role of realistic MC simulations has been very recently emphasized by the authors of Ref.~\cite{holler2020run}, who proposed a new simulation and analysis framework as an alternative to the  current way MC templates are obtained.  In~the existing paradigm, simulations are generated from pre-defined observation and instrument settings, such as the zenith of the observation or the configuration of the camera. Each simulation can be then seen as a grid point in the setting-parameters space. The~analyzer willing to use a ``run''\endnote{In IACTs a \textit{run} is generally referred to as a data taking (lasting roughly half an hour) performed on a given target under the same conditions throughout the entire observation.} performed with some given settings has to look for the adjacent grid points either by interpolating them or by taking the closest one. Instead in the \textit{run}-wise simulation approach described in Ref.~\cite{holler2020run}, simulations are generated on a run-by-run basis. In~this way observational conditions are fully taken into account, thus leading to more realistic MC simulations that reduce systematic uncertainties and improve the scientific output of the statistical~analysis.

\subsection{Multivariate~Analysis}
\label{Sec:Multivar}
To date, we described techniques for parametrizing an event detected by IACTs. These parameters are then used for inferring the energy, the~arrival direction of the primary gamma ray, and discriminating variables. The~latter are used to tell how likely the event can be associated with a gamma ray. The~usage of discriminating variables is quite simple: all the events with values of the parameter larger (or smaller, depending on the kind of variable) than a pre-defined threshold are retained and considered to be \textit{gamma-like} events, i.e.,~originating from a gamma ray. A~different approach that avoids cutting data by exploiting the full probability distribution function (PDF) of the discriminating variable will be discussed in Section~\ref{Sec:detection}. Once a discriminating variable is chosen and a fixed threshold is defined, the~separation (or discrimination) power can be obtained from the so-called $Q$~value
\begin{equation}
    Q = \frac{\epsilon_{\gamma}}{\sqrt{\epsilon_{h}}}.
\end{equation}
where $\epsilon_x$ is the efficiency of the selection procedure given by the fraction of events belonging to the population $x$ surviving the selection ($h$ stands for \textit{hadrons} which compose the background population). This classification problem becomes considerably more difficult when more than one parameter can be actually used for discriminating signal events from the background population.
Multivariate methods consist of combining several of the shower parameters into one single discriminating parameter.  The~main advantages of these approaches are~that
\begin{itemize}
    \item nonlinear correlations between the parameters are taken into account and
    \item those parameters with no discrimination power are ignored. 
\end{itemize}

\begin{figure}[H]
    
    \includegraphics[width=0.44\linewidth,height=0.33\linewidth]{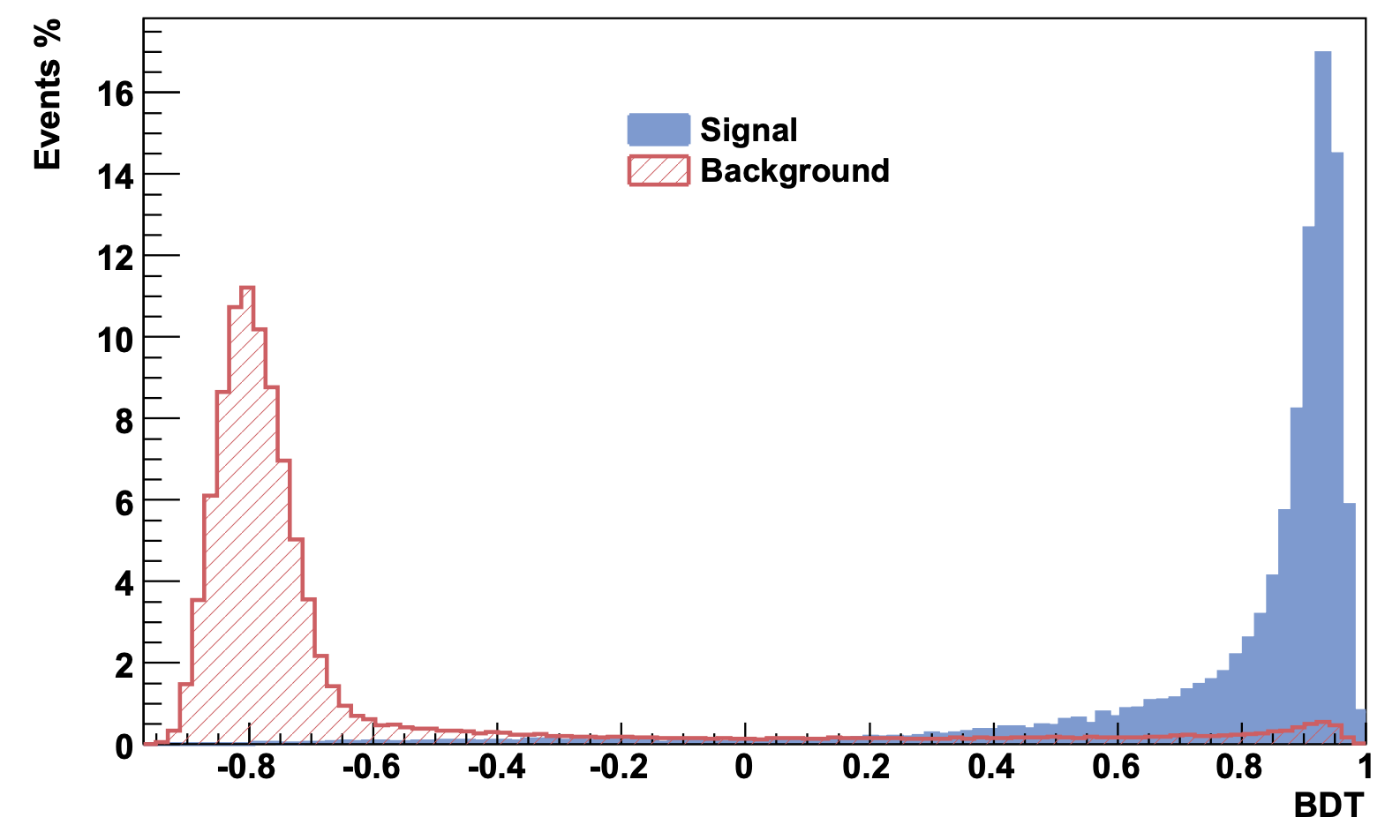}
        \includegraphics[width=0.54\linewidth,height=0.33\linewidth]{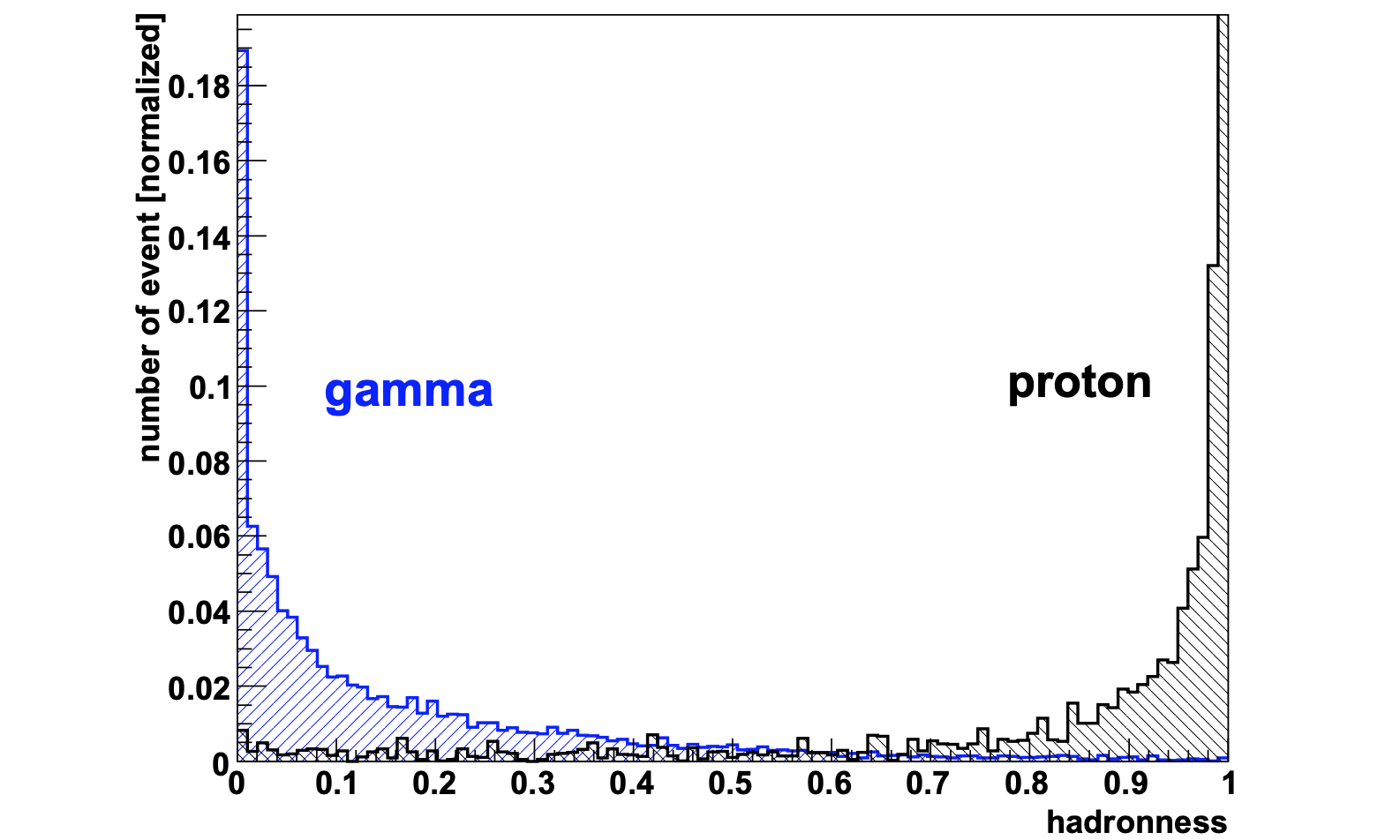}
    \caption{\textbf{Left panel}:  distribution for background events (hatched red) and simulated $\gamma$  (blue filled) of the discriminating variable given in output from the BDT method implemented by the HESS collaboration. Reprinted with permission from ref. ~\cite{fiasson2010optimization} Copyright 2010 Fiasson  et al. \textbf{Right panel}: distribution for background events (black) and simulated $\gamma$  (blue) of the discriminating variable (called \textit{hadronness}) given in output from the RF method implemented by the MAGIC collaboration. Reprinted with permission from Ref.~\cite{colin2009performance} Copyright 2009 Colin  et al.}
    \label{fig:DiscrimVar}
\end{figure}

A detailed review and comparison of different multivariate methods for event classification in IACTs can be found in Ref.~\cite{bock2004methods}. In~this section, we  provide a brief description of the currently most used methods, the~\textit{Boosted Decision Tree} (BDT) and the \mbox{\textit{Random Forest} (RF) \cite{breiman2017classification}}. The~BDT approach, implemented for the HESS~\cite{ohm2009gamma,fiasson2010optimization} and VERITAS~\cite{krause2017improved} telescopes, is a binary tree where events are sorted into small subsets by applying a series of cuts until a given condition is fulfilled. This condition might be given by requiring that the number of events in a leaf is smaller than a predefined value or that the signal over the background ratio in a leaf must be large enough.   The~term ``boosted'' refers to the fact that more than one individual decision trees are combined in a single classifier by performing a weighted average. The~boosting allows improving the stability of the technique
 with respect to fluctuations in the training sample and is able to
considerably enhance the performance of the  gamma/hadron separation compared to a single decision tree. Like the BDT approach, the~RF method, implemented for the MAGIC telescopes~\cite{albert2008implementation},  also relies on decision trees, which are built up and combined with some elements of random choice. As~for the BDT, training samples of the two classes of population (signal and background) are needed for constructing the decision trees. Once the classifier has been properly\endnote{On the one hand it is important to train the classifier to maximize the separation between the signal and background, and~on the other it is also crucial to avoid overtraining (also referred to as overfitting), i.e.,~avoiding the classifier to characterize statistical fluctuations from the training samples wrongly as true features of the event classes.} trained, the~algorithm can be  used to assign to each event a single discriminating variable whose distribution on a test gamma and hadron population can be seen in Figure~\ref{fig:DiscrimVar}. 

\unskip

\subsection{Deep Learning~Methods}
\label{Sec:CNNs}

The multivariate methods described in Section~\ref{Sec:Multivar} for discriminating the signal events from the background have shown a great capability in improving the sensitivity of IACTs. This effort has been recently pushed forward by Deep Learning (DL) \cite{goodfellow2016deep} techniques for object recognition in images. Such algorithms, which require more computational power, are getting more and more attention thanks to the improvements during recent decades in the usage of CPU and GPUs for matrix operations. When it comes to image processing, the~leading DL algorithm is Convolutional Neural Networks (CNNs) whose first application in the context of IACTs can be found in Ref.~\cite{feng2016analysis}, where a CNN was applied in the simple case of muon-ring events. This work served as a pathfinder for the application of CNNs for gamma/hadron separation from the raw recorded~images.

A CNN is made of many connected layers which in turn consist  of different nodes. The~first layer is the input image whose pixels represent its nodes.  The~inputs in a new layer  are convolved with kernels that have to be trained.  Each new layer is in general much smaller than the input one, and~allows identifying features in
the previous layer. Adding more and more layers, one aims to extract more and more complex features, which can be possibly used to identify those discriminating features  in the images that otherwise would not be considered in other event-classifier algorithms. For~a more detailed review and description of DL and CNNs algorithms, we refer the reader to Refs.~\cite{goodfellow2016deep,lecun2015deep} and references therein. The~training process in CNN can be
computationally demanding, due mainly to the very large  number of parameters.  The~main advantages of DL relative to previous event reconstruction methods in IACT  is that CNNs do not need the image parameters (such as the Hillas ones), and~therefore all the features contained in the image are fully exploited, while they might get lost or suppressed during the parametrization. Recent applications of CNNs in the image processing of IACTs data can be found in Refs.~\cite{shilon2019application,mangano2018extracting,NietoCasta,Holch},  where the algorithm was also implemented for the energy and arrival direction estimation of the~gamma rays.

\section{Detection Significance and Background~Modeling }
 \label{Sec:detection}
The final result of the statistical analysis described in Section~\ref{Sec:EventRecon} is a list containing useful information about the candidate gamma ray events, such as their estimated energy and arrival direction. Neglecting any background contamination, the~total number $n$ of events in such a list would be a random variable distributed according to the Poisson PMF
\begin{equation}
    \mathcal{P}( n | s ) =  \frac{s^{n}}{n!} e^{-s},
    \label{Eq:Poisson}
\end{equation}
where $s$ is the expected number of signal counts.
The problem is  that the majority of the observed events are actually generated by hadronic cosmic rays, while only a small fraction (which for the case of a bright source such as the Crab Nebula is of the order of $10^{-3}$) can be associated with gamma rays. By~applying a signal extraction selection on the data based on a discrimination variable, it is possible to reduce the background by a factor of 100 or more, thus increasing the signal-to-noise ratio close to 1 for a bright source.  In~order to infer the gamma ray flux from the resulting event list it is essential to first estimate the remaining background contamination.  We can consider three different scenarios (which are the topic of Sections~\ref{Subsec:b0}--\ref{Subsec:b2}, respectively) where the expected background count $b$ in the target region~is:
\begin{itemize}
    \item  zero or negligible relative to the source counts,
    \item  known precisely,
    \item  estimated from an OFF measurement.
\end{itemize}

The latter case is the most common one and requires the definition of two regions: a region of interest (ROI), also referred to as target, test or ON region, and~a background control region, called OFF region. The~ON and OFF regions provide, respectively, independent  $N_{on}$ and $N_{off}$ counts, where the latter is  ideally void of any signal event. A~ normalization factor $\alpha$ is introduced to account for differences (such as the acceptance and the exposure time) between the ON  and OFF region. It can be defined as
\begin{equation}
    \alpha = \frac{\int_{ON} A(x,t) dx dt}{\int_{OFF} A(x,t) dx dt},
\end{equation}
where $A(x,t)$ is the instrument acceptance, which is a function of the observation time $t$ and of all  observational parameters (such as the FoV position or the zenith angle) here denoted for simplicity with $x$. The~goal of the background modeling analysis is to provide the values of $\alpha$ and $N_{off}$ that are then used for estimating the signal $s$ along with the detection~significance. 

\subsection{The Background Is Zero or~Negligible} 
\label{Subsec:b0}
Although very rare, in~some analyses the background $b$ in the  ON region may be assumed to be zero or negligible relative to the signal $s$. Given the simplicity of this case, it is worth dwelling on it, discussing with examples the statistical conclusions that can be drawn from a measurement using the frequentist and Bayesian approach.
    In this scenario the  likelihood function is trivially
\begin{equation}
        \mathcal{L}(s) = \mathcal{P}( N_{on} | s ), 
    \end{equation}
    where $\mathcal{P}$ is the Poisson distribution (see Equation~(\ref{Eq:Poisson})) with observed and expected counts $N_{on}$ and $s$, respectively.
    Using the likelihood ratio defined in Equation~(\ref{Eq:LKratio}) as a statistic, and~taking into account that $\hat{s}=N_{on}$ is the value of $s$ that maximizes the likelihood, we get for any $s>0$ the following statistic
\begin{equation}
        \mathcal{S} = 2 \cdot (s - N_{on}) - 2 N_{on} \cdot \left( \log s - \log N_{on}   \right).
        \label{Eq:stat_b0}
    \end{equation}

\begin{figure}[H]
    
    \includegraphics[width=0.99\linewidth]{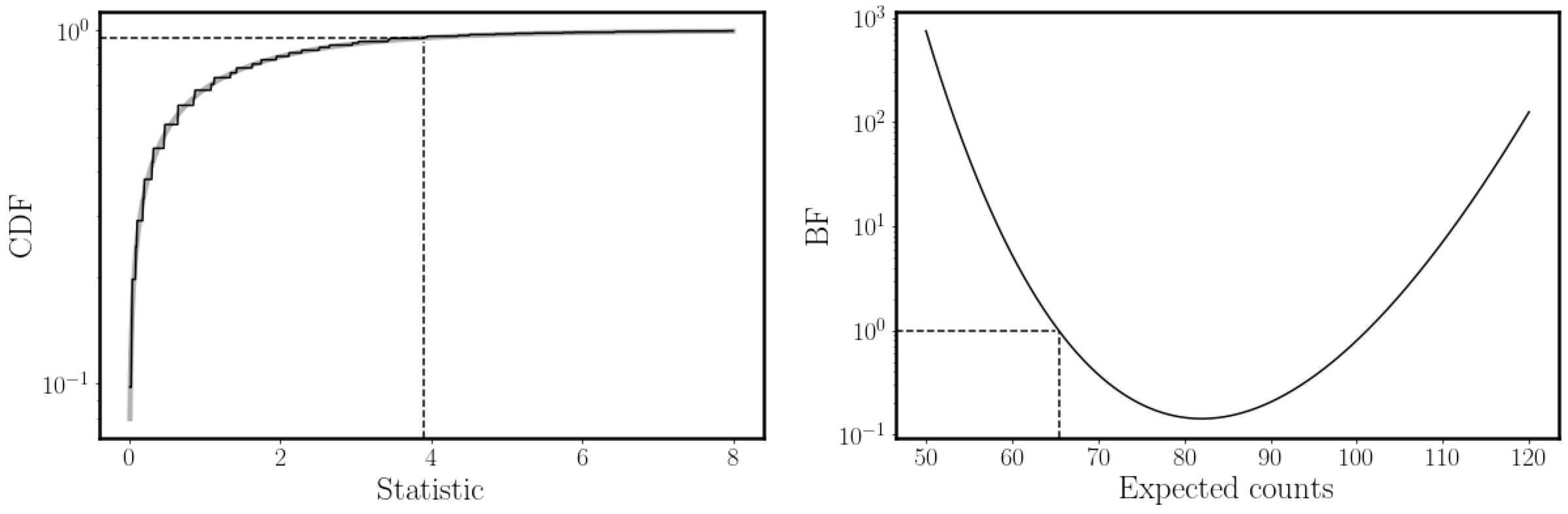}
    \caption{A comparison between the frequentist (\textbf{left panel}) and Bayesian (\textbf{right panel}) conclusion from the experiment result $N_{on} = 82$ on the hypothesis that the gamma ray expected counts is 65.4 and no background is present. Left panel: in black the cumulative distribution function (CDF) of the statistic defined in Equation~(\ref{Eq:stat_b0}) from $10^6$ simulations assuming $s=65.4$, while in grey the expected CDF of a $\chi^2$ random variable. The~step shape of the CDF of the statistic is due to the discrete nature of the Poisson distribution. Dashed lines show the point in which the statistic is 3.9 and the CDF is 0.952. Right panel:  evolution of the BF defined in Equation~(\ref{Eq:BF}) as a function of the expected counts $s_2$, using $s_1 = 65.4$  and $N_{on}=82$. Dashed lines show the point in which the expected counts are 65.4 and by definition the BF is~1. }
    \label{fig:CDF_BF}
\end{figure}

Such a statistic, known in the literature as the Cash statistic or C-statistic~\cite{1979ApJ...228..939C}, has a straightforward meaning: if we measured $N_{on}$ counts in the ON region and assumed that the true signal rate is $s$, then the value obtained from $\mathcal{S}$ according to Wilks' theorem is a random variable that follows a $\chi^2$ distribution with 1 degree of freedom. This can be checked by performing MC simulations as shown in the left plot of Figure~\ref{fig:CDF_BF}. The~smaller the true value of $s$ the more difficult it is to find an exact distribution for the statistic $\mathcal{S}$ which, due to the discreteness of the Poisson distribution, cannot be assumed anymore to be a  $\chi^2$ variable. For~small expected signal counts $s$, it is therefore necessary to get the CDF of $\mathcal{S}$ from MC simulations. 

We can be interested, for~instance, in~the {hypothesis} $\mathcal{H}_0$: ``the number of expected signal events (for a given temporal and energetic bin and surface area) is $s=65.4$''. After~having performed the experiment, we obtain from the measurement $N_{on} =82$ events. In~this scenario, by computing the statistic in Equation~(\ref{Eq:stat_b0}) we get $\mathcal{S} = 3.9$. 
If $\mathcal{H}_0$ was true we would have observed a  value of the statistic equal or greater than 3.9 only $4.8 \%$ of the time (see the left panel of Figure~\ref{fig:CDF_BF}).   The~conclusion of the frequentist approach is therefore that $\mathcal{H}_0$ can be excluded with a $95.2 \%$ CL or in other words with a 1.98 $\sigma$ significance. 
The latter is obtained by expressing the CL in multiples of the standard deviation $\sigma$ of a normal distribution via the inverse of the error function\endnote{One can check that by computing $\sqrt{ \mathcal{S}}$ one would get the same value of Equation~(\ref{Eq:erf}). That is because $\mathcal{S}$ is a $\chi^2$ random variable.}:
\begin{equation}
     \sqrt{2} \; \text{erf}^{-1} \left( \text{CL} \right) .
     \label{Eq:erf}
\end{equation}

The aim of the Bayesian approach is instead to provide a probabilistic statement about $s$, which is not fixed as in the frequentist approach. By~applying the Bayes theorem,  we get that the PDF of $s$ is (up to a normalization factor)
\begin{equation}
    p(s | N_{on} ) \propto  \mathcal{P}( N_{on} | s ) \cdot p(s),
    \label{Eq:bayes_s}
\end{equation}
where $p(s)$ is the prior PDF of $s$, which encloses the prior knowledge the analyzer has on the source's signal. In~the Bayesian context we can compare two hypotheses $s=s_1$ and $s=s_2$ as follows
\begin{equation}
    \frac{p( s_1 | N_{on} )}{p( s_2 | N_{on} )} = \text{BF} \cdot \frac{p( s_1  )}{p( s_2  )}, 
\end{equation}
where
\begin{equation}
    \text{BF} = \left( \frac{s_1}{s_2} \right)^{N_{on}} e^{-s_1 + s_2}.
    \label{Eq:BF}
\end{equation}

The evolution of the BF as a function of the expected counts $s_2$, using $s_1 = 65.4$ (which is our hypothesis $\mathcal{H}_0$ of interest) and the experiment result $N_{on}=82$ is shown in the right panel of Figure~\ref{fig:CDF_BF}. It is worth noticing that the BF is connected to the statistic in Equation~(\ref{Eq:stat_b0})~via
\begin{equation}
    - 2 \log BF = \mathcal{S}, 
\end{equation}
in which $s_1 = s$ and $s_2 = N_{on}$. Lastly, it can be shown that assuming a uniform prior\endnote{See for instance Ref.~\cite{d1998jeffreys} for a review of the problem regarding the choice of the priors.} the PDF of $s$ in Equation~(\ref{Eq:bayes_s}) is
\begin{equation}
    p(s | N_{on} ) = \mathcal{P}( N_{on} | s )  = \frac{s^{N_{on}}}{N_{on}!} e^{-s}.
\end{equation}

\subsection{The Background Is Known~Precisely} 
\label{Subsec:b1}

This is the case in which we know from theoretical or experimental considerations the true value $\bar{b}$ of the background. This happens for instance in the \textit{field-of-view}-background model, where the entire FoV (excluding positions where $\gamma$-ray events are expected) is used for modeling the background. Since the OFF region is composed by the entire FoV and the ON region by a small portion of it, we have $\alpha~{\ll}~1$ . Therefore the Poissonian fluctuations of the background contamination in the ON region   can be neglected, being given by 
\begin{equation}
    \sigma( \alpha N_{off} ) = \alpha \sqrt{ N_{off}}.
\end{equation}

Indeed the detection significance of the ``known'' and ``unknown''  background cases coincide for $\alpha~{\ll}~1$ (see Section~\ref{Subsec:b2}).
Thus, in~the \textit{field-of-view}-background model we can make the assumption of knowing precisely the background, given by the product $ \alpha N_{off}$.

The likelihood function is
\begin{equation}
    \mathcal{L}(s) = \mathcal{P}( N_{on} | s + \bar{b} ),
\end{equation}
which reaches its maximum value for $\hat{s}= N_{on} - \bar{b}$.  The~statistic is obtained from the Cash one defined in Equation~(\ref{Eq:stat_b0}) in which $s$ has been substituted with $s + \bar{b}$. An~important difference between the previous case, in~which the background was assumed to be zero, is that now the statistic is also defined for the hypothesis $s=0$.  For~this no-source hypothesis it is common to slightly modify the Cash statistic by taking the square root of it and by introducing a sign that is arbitrarily chosen to be positive when the excess $N_{on} - \bar{b}$ is positive, yielding
\begin{equation}
        \mathcal{S} = \pm \sqrt{ 2 \cdot (\bar{b} - N_{on}) - 2 N_{on} \cdot \left( \log \bar{b} - \log N_{on}   \right) }.
        \label{Eq:stat_b1}
    \end{equation}

\begin{figure}[H]
    
    \includegraphics[width=0.99\linewidth]{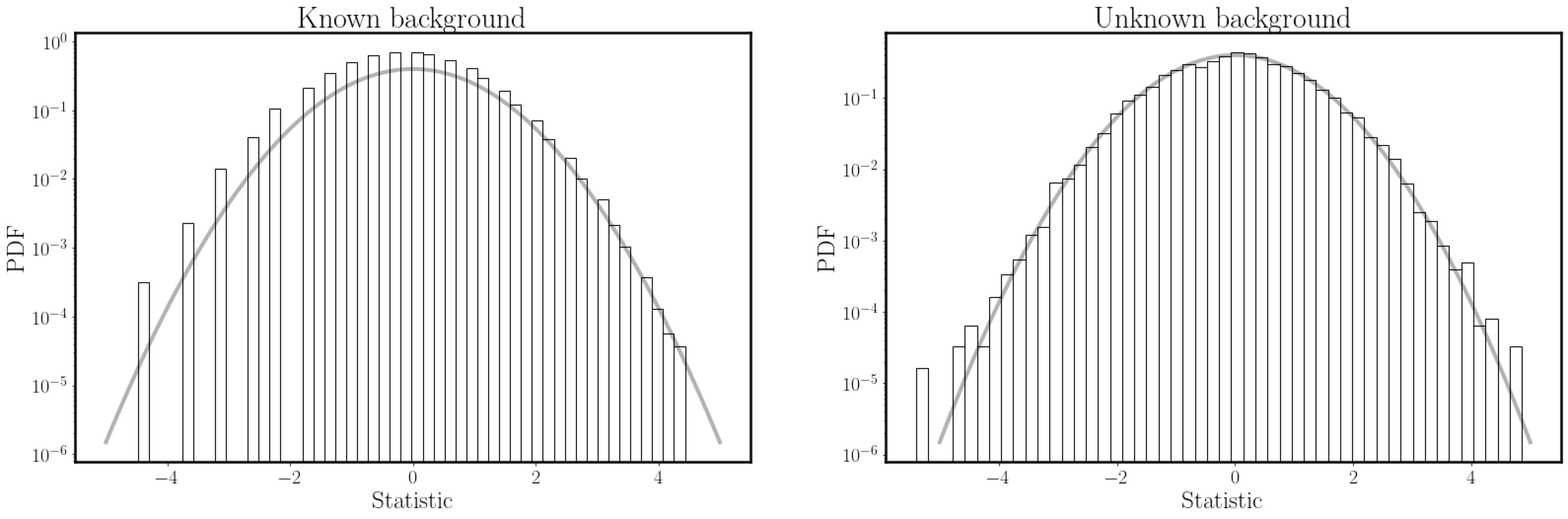}
    \caption{Distribution of the Cash statistic in Equation~(\ref{Eq:stat_b1}) (\textbf{left panel}) and the Li\&Ma statistic in Equation~(\ref{Eq:S_LiMa}) (\textbf{right panel}) from measurements in which the background is known precisely to be $\bar{b}=10$ or must be estimated from the OFF counts $N_{off}$ with $\alpha =1$, respectively. In~both simulations, the true values of $s$ and $b$ are 0 and 10, respectively. In~both plots, the PDF of a normal distribution with mean zero and variance 1 is shown in grey as reference. In~both cases $10^{6}$ simulations were performed. The~distribution in the left panel looks to be less populated due to the fact that having the background fixed to 10 (and not estimated from an OFF measurement) limits the number of possible outcomes of the~statistic.}
    \label{fig:PDF_LiMa}
\end{figure}

In this way, for~large enough counts ($N_{on} \gtrsim 10 $) the statistic is a random variable that follows a normal distribution with mean zero and variance 1 (as shown in the left panel of Figure~\ref{fig:PDF_LiMa}). This allows immediately converting the output of $\mathcal{S} $ in a significance level. If~for instance we assume $\bar{b} = 10$ and we observe $N_{on}=21$ events, then $\mathcal{S} = 3.0$ which means that the no-source hypothesis can be excluded with a significance of 3 $\sigma$. 

\unskip

\subsection{The Background Is Estimated from an OFF~Measurement}
\label{Subsec:b2}

Let us consider the most common scenario, in~which we do not know the background $b$ and therefore need  to estimate it by performing OFF measurements, supposedly void of any signal. Such OFF measurements can be performed following one of the below~procedures:
\begin{itemize}
    \item  \textit{On-Off background}: the OFF counts are taken from (usually consecutive) observations  made under identical conditions,  meaning that $\alpha$ is simply given by $t_{on}/t_{off}$ with $t_{on}$ and $t_{off}$ the exposure time for the ON and OFF observation, respectively. The~main advantage of this method is that no assumption is required for the acceptance, except~that it is the same for the ON and OFF regions. The~drawback of this approach is that dedicated OFF observations are needed, thus ``stealing'' time from the on-source~ones.
    
    \item \textit{Wobble or reflected-region background}: the OFF counts are taken from regions located, on~a run-by-run basis, at~identical distances  from the center of the field of view. Each of the OFF regions is obtained by reflecting the ON region  with respect to the FoV center. This is the reason this method is called the  \textit{reflected-region} method. If~we have $n$ OFF regions then $\alpha$ is equal to $1/n$. This technique was originally applied to \textit{wobble} observations~\cite{aharonian2001evidence} and was later on used also in other observation modes. 
    \item \textit{Ring background}: the OFF counts are taken from a ring around the ROI or around the center of the field of view. 
    \item \textit{Template background}: the OFF counts are given by those events that have been discarded in the signal extraction selection based on a discriminating variable. In~this method, first developed for the \textit{HEGRA} experiment~\cite{rowell2003new} and more recently refined for HESS~\cite{fernandes2014new}, the~discarded events are used to template the background. 
\end{itemize}

For a more detailed review of the background modeling and comparison of the different methods see Ref.~\cite{berge2007background}.

From one of the above-mentioned procedures, once we obtained  the value of $\alpha$ and $N_{off}$, the~inference analysis on $s$ is performed using  the following likelihood:
\begin{align}
\mathcal{L}(  s,b ) &  = \, \mathcal{P}( N_{on}  \; | \;  s+ \alpha b) \cdot \mathcal{P}(  N_{off} \; | \;  b)  = \nonumber \\
 & =   \frac{(s + \alpha b)^{N_{on}}}{N_{on}!} e^{- (s +\alpha b) } \cdot \frac{ b^{N_{off}}}{N_{off}!} e^{-  b }.
 \label{Eq:Likelihood_b2}
\end{align}

\begin{figure}[H]
    
    \includegraphics[width=0.99\linewidth]{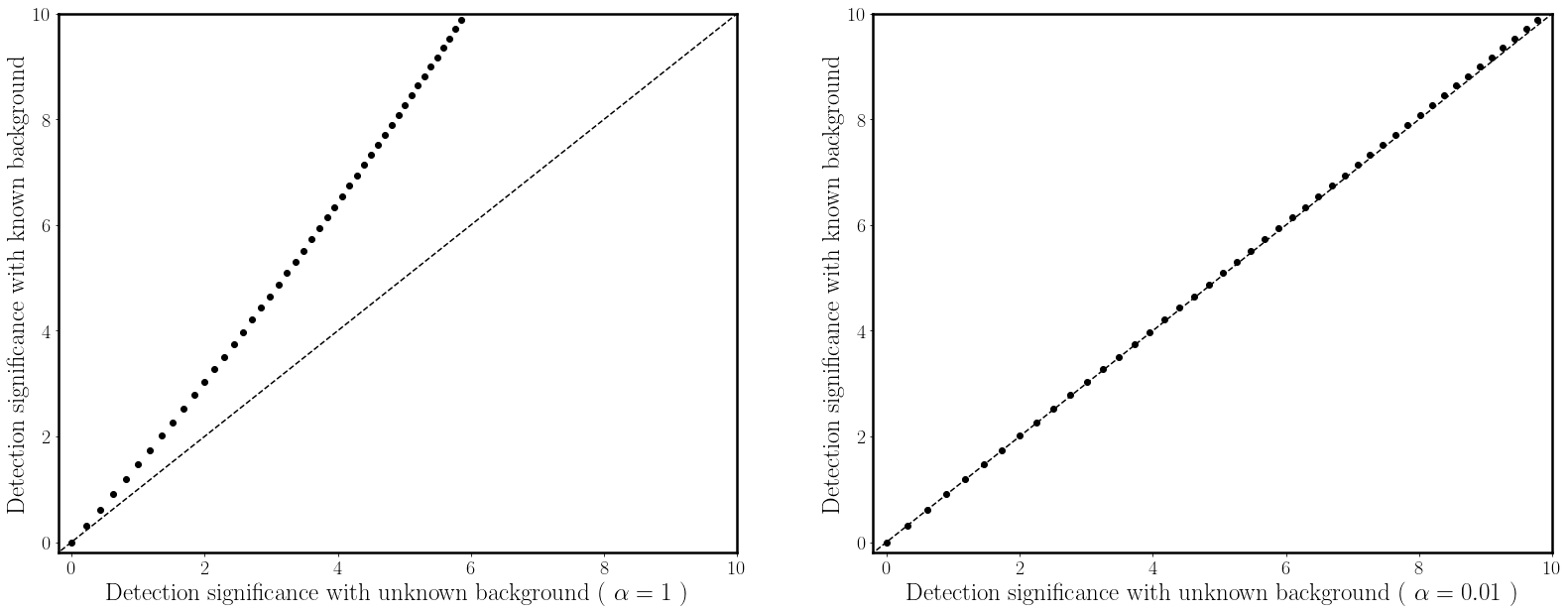}
    \caption{Comparison between  the Li\&Ma statistic in Equation~(\ref{Eq:S_LiMa}) (x-axis) and the Cash statistic in Equation~(\ref{Eq:stat_b1}) (y-axis). In~both plots, each point shows the significance for a different $N_{on}$ ranging from 10 (where the significance is zero in both cases) to 56. For~the Cash formula $\bar{b}$ is fixed to 10 in both plots, while for the Li\&Ma formula  $N_{off}$ is 10 with $\alpha=1$ in the left plot, and~$N_{off}=1000$ with $\alpha=0.01$ in the right plot.  As~a reference  the equation $y=x$ (dashed line) is shown. One can see that the Li\&Ma  statistic converges to the Cash one when $\alpha~{\ll}~1$. }
    \label{fig:Cash_lima}
\end{figure}

  We are not directly interested in knowing the background, which is therefore a nuisance parameter. Thus, in~the frequentist approach we have to profile the likelihood  (see \mbox{Equation~(\ref{Eq:Lkl_profile})}) by fixing $b$ to the value $\hat{\hat{b}}$ that maximizes $\mathcal{L}$ for a given  $s$ (see for instance Ref.~\cite{rolke2001confidence} for a derivation of $\hat{\hat{b}}$) , i.e.,
\begin{equation}
\hat{\hat{b}} = \frac{N + \sqrt{N^2 + 4(1+ 1/\alpha)s N_{off}} }{2(1+ \alpha)},
\label{Eq:hathatb}
\end{equation}
with $N \equiv N_{on} + N_{off} - (1+ 1/\alpha)s$. Performing as usual the logarithm of the likelihood ratio  we have
\begin{align}
     \mathcal{S} = -2 \log \frac{\mathcal{L}(s,\hat{\hat{b}})}{\mathcal{L}(\hat{s},\hat{b})} = &
    \;  2 \left[   N_{on} \log \left( \frac{N_{on}}{s+ \alpha \hat{\hat{b}} } \right) + 
     N_{off} \log \left( \frac{N_{off}}{ \hat{\hat{b}} } \right) +  \right. \nonumber \\
     & \left. + \;  s+ (1+ \alpha) \hat{\hat{b}} - N_{on} - N_{off} \right],
     \label{Eq:Lklratio_b2}
\end{align}
  where  $\hat{s} = N_{on}- \alpha N_{off}$ and $\hat{b} = N_{off}$ are the values\endnote{Indeed one can check that  Equation~(\ref{Eq:hathatb})  yields $\hat{\hat{b}} = N_{off}$ when $s = N_{on}- \alpha N_{off} $ .} that maximizes the likelihood in Equation~(\ref{Eq:Likelihood_b2}), while $\hat{\hat{b}}$ is given in Equation~(\ref{Eq:hathatb}) and maximizes $\mathcal{L}$ for a given  $s$ . The~statistic in Equation~(\ref{Eq:Lklratio_b2}) depends only on  the free parameter $s$  and according to Wilks' theorem it follows a $\chi^2$ distribution with 1 degree of freedom. It can then be used for  hypothesis testing, in~particular the ``$s=0$'' hypothesis\endnote{If $s=0$ then $\hat{\hat{b}} = (N_{on} + N_{off})/(1+\alpha)$ and the term $(1+ \alpha) \hat{\hat{b}} - N_{on} - N_{off}$ in Equation~(\ref{Eq:Lklratio_b2}) vanishes.} from which we can obtain the detection significance. Similarly to Equation~(\ref{Eq:stat_b1}), we can take the square root of Equation~(\ref{Eq:Lklratio_b2}) and set $s=0$, yielding the statistic
\begin{align}
   S = \pm \sqrt{2}  \left[  
   N_{on} \log\left( \frac{1 }{\alpha} \frac{(\alpha +1) N_{on}}{N_{on} + N_{off}} \right)  +
    N_{off} \log\left(  \frac{(\alpha +1 ) N_{off}}{N_{on} + N_{off}} \right) \right]^{1/2},
    \label{Eq:S_LiMa}
\end{align}
where the sign is arbitrarily chosen to be positive when the excess $N_{on} - \alpha N_{off}$ is positive. This expression is the well-known ``Li\&Ma'' \cite{Li_Ma} formula for computing the detection significance  in ON/OFF measurements.  As~shown in the right panel of Figure~\ref{fig:PDF_LiMa}  for large enough counts ($N_{on}, N_{off} \gtrsim 10 $) the statistic in Equation~(\ref{Eq:S_LiMa}) distributes according to a normal distribution with mean zero and variance 1. We can again consider the example in which  $N_{on} = 21$ counts have been observed in the ON region, but~instead of assuming a known background $\bar{b}=10$, our background is instead estimated from the OFF measurement $N_{off} = 10$ with $\alpha = 1$. Using the statistic in Equation~(\ref{Eq:S_LiMa}) we get a detection significance of 2 $\sigma$, which is smaller than the 3  $\sigma$ obtained from the Cash statistic where the background is assumed to be known precisely.  A~comparison of different values of $N_{on}$ between the Cash (Equation~(\ref{Eq:stat_b1})) and Li\&Ma (Equation~(\ref{Eq:S_LiMa})) statistic  is shown in Figure~\ref{fig:Cash_lima}, where one can see that the former becomes  bigger than the latter as more events are observed in the ON region. 
This is due to the fact that the Li\&Ma statistics account for the Poissonian fluctuations in the observed counts $N_{off}$. These fluctuations make an association with the gamma ray excess with the source signal less likely.
 When $\alpha~{\ll}~1$ the two statistics give the same~result.

The Li\&Ma expression in Equation~(\ref{Eq:S_LiMa}) based on the likelihood ratio  is not the only statistic used for rejecting the background-only hypothesis in Poisson counting experiments. One can find in the literature the so-called ``signal-to-noise ratio''
\begin{equation}
    \mathcal{S} = \frac{N_{on}-\alpha N_{off}}{\sqrt{N_{on} + \alpha^2 N_{off}} },
\end{equation}
which  has the disadvantages of following a normal distribution only for values of $\alpha$ close to 1 and for large enough counts~\cite{Li_Ma}. Another approach is to compute the \emph{p}-value from the observed $N_{on}$ counts, i.e.,~the probability of observing a total count bigger than $N_{on}$ under the assumption of the only-background hypothesis. If~we ignore uncertainties in the background we have
\begin{equation}
    \text{\emph{p}-value} = \sum_{n = N_{on}}^{\infty} \mathcal{P}(n | \alpha b) = \frac{\Gamma( N_{on}, 0, \alpha b) }{ \Gamma(N_{on})},
\end{equation}
written in terms of the incomplete gamma  function $\Gamma$. When we want to include the fact that the background is estimated from an OFF measurement, it is convenient to introduce the variable $N_{tot} = N_{on} + N_{off}$, and~it can be shown~\cite{cousins2008evaluation} that the observed quantity  $N_{on}$ follows (for a given $N_{tot}$) a binomial distribution $\mathcal{B}$ with success probability $\rho \equiv \alpha / (1+ \alpha)$ and total numer of attempts $N_{tot}$. The~\emph{p}-value is
\begin{equation}
    \text{\emph{p}-value} = \sum_{n = N_{on}}^{\infty} \mathcal{B}(n | N_{tot}, \rho ) = \frac{ \beta( \rho, N_{on}, N_{off} +1)}{\beta( N_{on}, N_{off} +1)},
\end{equation}
with $\beta$ the incomplete and complete beta functions (distinguished by the number of arguments).
Finally the statistic is defined from the above \emph{p}-values using
\begin{equation}
    \mathcal{S} = \sqrt{2} \; \text{erf}^{-1} ( 1 - 2 \cdot \text{ \emph{p}-value} ).
\end{equation}

A review and comparison of these statistics applied to the ON/OFF measurement can be found in Refs.~\cite{Li_Ma, linnemann2003measures,cousins2008evaluation, vianello2018significance}. 
Finally, it is worth mentioning that different extensions and refined versions of the Li\&Ma expression in Equation~(\ref{Eq:S_LiMa}) were introduced, each one with its application to a particular problem. The~problem of including PSF\endnote{PSF stands for Point Spread Function. See Section~\ref{Sec:Flux} for its definition.} information in the likelihood is addressed in Refs.~\cite{klepser2012generalized, nievas2016extending, klepser2017optimal} , while the problem of including the prior knowledge of the source light curve is considered in Ref.~\cite{weiner2015new} . Assessing the role in the detection significance of systematic uncertainties, especially those rising from the normalization factor $\alpha$, is performed in the studies of Refs.~\cite{dickinson2013handling, spengler2015significance, vianello2018significance}.  
Following the prescriptions of probability theory, in~the Bayesian approach, the signal $s$ is estimated by defining its PDF in which the nuisance parameter $b$ has been marginalized:
\begin{align}
p( s & \; | \;  N_{on},  N_{off}; \alpha)   \propto
  \int_{0}^{\infty} db \; \mathcal{P}(  N_{on} | s + \alpha b )    \mathcal{P}(  N_{off} |  b )   p(b)\,p(s).
\label{Eq:Integral_b}
\end{align}

\begin{figure}[H]
    
    \includegraphics[width=0.95\linewidth]{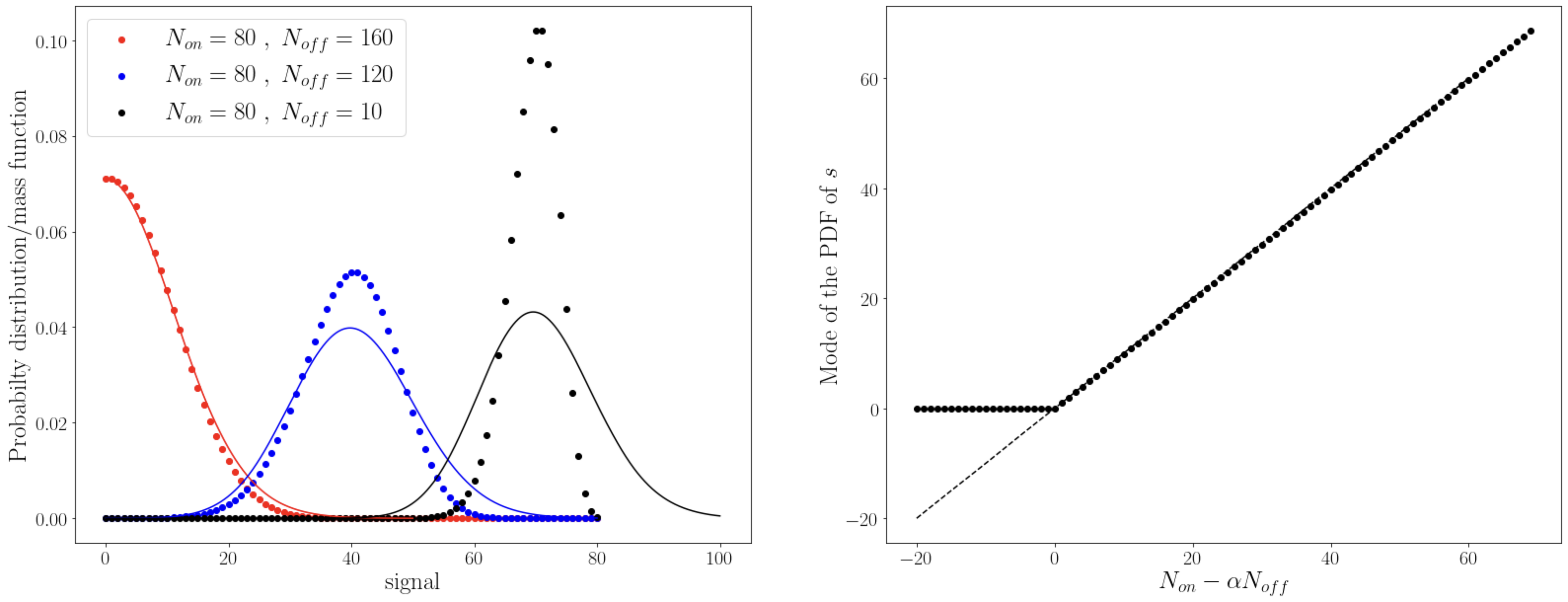}
    \caption{\textbf{Left panel}: Points show the PMF defined in Equation~(\ref{Eq:PDF_Nexc}) of the number of signal events $N_s$, while lines show the PDF defined in Equation~(\ref{Eq:PDF_signal}) of the expected signal counts $s$. Different colors are used to distinguish the different counts $N_{off}$ (160 in red, 120 in blue and 10 in black), while $N_{on}$ and $\alpha$ are fixed to 80 and 0.5, respectively. \textbf{Right panel}: The mode of the PDF defined in Equation~(\ref{Eq:PDF_signal}) as a function of the excess $N_{on} - \alpha N_{off}$. As~a dashed line the equation $x=y$ is shown for~reference. }
    \label{fig:PDF_PMF_s}
\end{figure}

Assuming flat priors $p(s)$ and $p(b)$ (with $s>0$ and $b>0$) it can be shown~\cite{Loredo} that the integral in Equation~(\ref{Eq:Integral_b}) is
\begin{equation}
    p( s  \; | \;  N_{on},  N_{off}; \alpha)   \propto \sum_{N_s = 0}^{N_{on}}  \frac{ (N_{on} + N_{off} - N_s)! }{(1+1/\alpha)^{-N_s}  (N_{on} - N_s)! }   \cdot  \frac{s^{N_s}}{N_s !} e^{-s},
    \label{Eq:PDF_signal}
\end{equation}
where $N_s$ is a bound variable whose physical meaning will be clear soon. Thus, the~PDF of the expected signal counts $s$ is given by a weighted sum of Poisson distributions with observed counts ranging from 0 to $N_{on}$. One can recognize (see Refs.~\cite{Loredo, d2021signal} for a detailed explanation) in Equation~(\ref{Eq:PDF_signal}) a marginalization over the variable $N_s$. The~weights in the sum of Equation~(\ref{Eq:PDF_signal}) are indeed the PMF of the number of signal events $N_s$ in the ON~region:
\begin{equation}
p(N_s \; | \; N_{on}, N_{off}; \alpha) \propto  \frac{ (N_{on} + N_{off} - N_s)! }{(1+1/\alpha)^{-N_s}  (N_{on} - N_s)! }\,.
\label{Eq:PDF_Nexc}
\end{equation}
In the left plot of Figure~\ref{fig:PDF_PMF_s} the PDF of $s$ and the PMF of $N_s$ from Equations~(\ref{Eq:PDF_signal}) and (\ref{Eq:PDF_Nexc}), respectively, are shown.
The best estimate of $s$ can be then obtained from the mode of the PDF in Equation~(\ref{Eq:PDF_signal}). The~evolution of the Bayesian estimation of $s$ as a function of the excess $N_{on} - \alpha N_{off}$ can be found in the right plot of Figure~\ref{fig:PDF_PMF_s}.
We can now compare the two hypotheses $\mathcal{H}_{s+b}$ and $\mathcal{H}_{b}$, respectively, ``the observed counts in the ON region are produced by the source and background'' and  ``the observed counts in the ON region are only produced by the background''. The~BF is (see Ref.~\cite{gregory2005bayesian}) in this case (assuming again flat priors for $s$ and $b$)
\begin{equation}
\text{BF} = \frac{p(N_{on} | N_{off} , \alpha, \mathcal{H}_{s+b}) }{p(N_{on} | N_{off} ; \alpha, \mathcal{H}_{b} )} =  \frac{1}{s_{max}} \frac{N_{on}!}{(N_{on} + N_{off}  )!} \cdot \sum_{N_s = 0}^{N_{on}}  \frac{ (N_{on} + N_{off} - N_s)! }{(1+1/\alpha)^{-N_s}  (N_{on} - N_s)! },
\end{equation}
where $s_{max}$ is the maximum prior value of $s$, i.e.,~$p(s) = 1/s_{max}$.
From the above expression, one can then compute the odds of $\mathcal{H}_{s+b}$ following
\begin{equation}
 o( \mathcal{H}_{s+b} |  N_{on}, N_{off} , \alpha) = \text{BF} \cdot o( \mathcal{H}_{s+b}) ,
\end{equation}
with
\begin{equation}
o( \mathcal{H}_{s+b})  = \frac{p(\mathcal{H}_{s+b}) }{1-p(\mathcal{H}_{s+b}) } = \frac{p(\mathcal{H}_{s+b}) }{p(\mathcal{H}_{b}) } ,
\end{equation}
and $p(\mathcal{H}_{s+b})$ and $p(\mathcal{H}_{b})$ the priors of the two competing hypothesis $\mathcal{H}_{s+b} $ and $\mathcal{H}_{b} $, respectively. The~above odds can be expressed in a ``frequentist-fashion'' way by converting the posterior probability of $\mathcal{H}_{b}$ in a significance value using the inverse error function:
\begin{equation}
 \mathcal{S} = \sqrt{2} \cdot \text{erf}^{-1} \left( 1-p(\mathcal{H}_{b} |N_{on}, N_{off} , \alpha)   \right)
\end{equation}
as shown for instance in Ref.~\cite{knoetig2014signal} where both constant and Jeffreys’s~\cite{jeffreys1998theory} priors are assumed and a comparison with the Li\&Ma significance (see Equation~(\ref{Eq:S_LiMa})) is shown. More recently this effort has been pushed forward in Ref.~\cite{casadei2014objective}, where an objective Bayesian solution is proposed and compared to the result of Ref.~\cite{knoetig2014signal}. The~main advantage of these Bayesian solutions is that there are no restrictions in the number of counts $N_{on}$ and $N_{off}$, while the frequentist ones holds only when the counts are large enough. Yet, it is important to not confuse the two approaches, since they aim at finding the solution of two different problems: studying the long-run performance of a statistic in the frequentist approach and deriving the probability of hypotheses in the Bayesian~approach. 

Lastly, it is worth mentioning that it is possible~\cite{d2021signal} to extend the PDF of $s$ in \mbox{Equation~(\ref{Eq:PDF_signal})} by including the information on how the discriminating variables distribute for a signal or background population. The~authors of Ref.~\cite{d2021signal} showed that by performing such extension not only can one avoid discarding data based on a discrimination variable (which inevitably discards also part of the signal events) but one can also increase the resolution of the signal~estimation.

\subsection{Bounds, Confidence and Credible~Intervals }
We have shown so far how, given the number of events observed in the ROI, one can estimate the source signal $s$ and its significance. However, the~statistical analysis would be incomplete without also reporting lower and upper bounds on the inferred parameters. In~the former case they are referred to in the literature as lower limits (LLs), while in the latter as  upper limits (ULs), with~the interpretation that values of the parameters  below the LL or above the UL are more unlikely to be true.  They are particularly useful when the detection is not significant, for~instance when the source is too dim, and~therefore one would like to provide an UL on the strength of the signal $s$.

\begin{figure}[H]
    
    \includegraphics[width=0.99\linewidth]{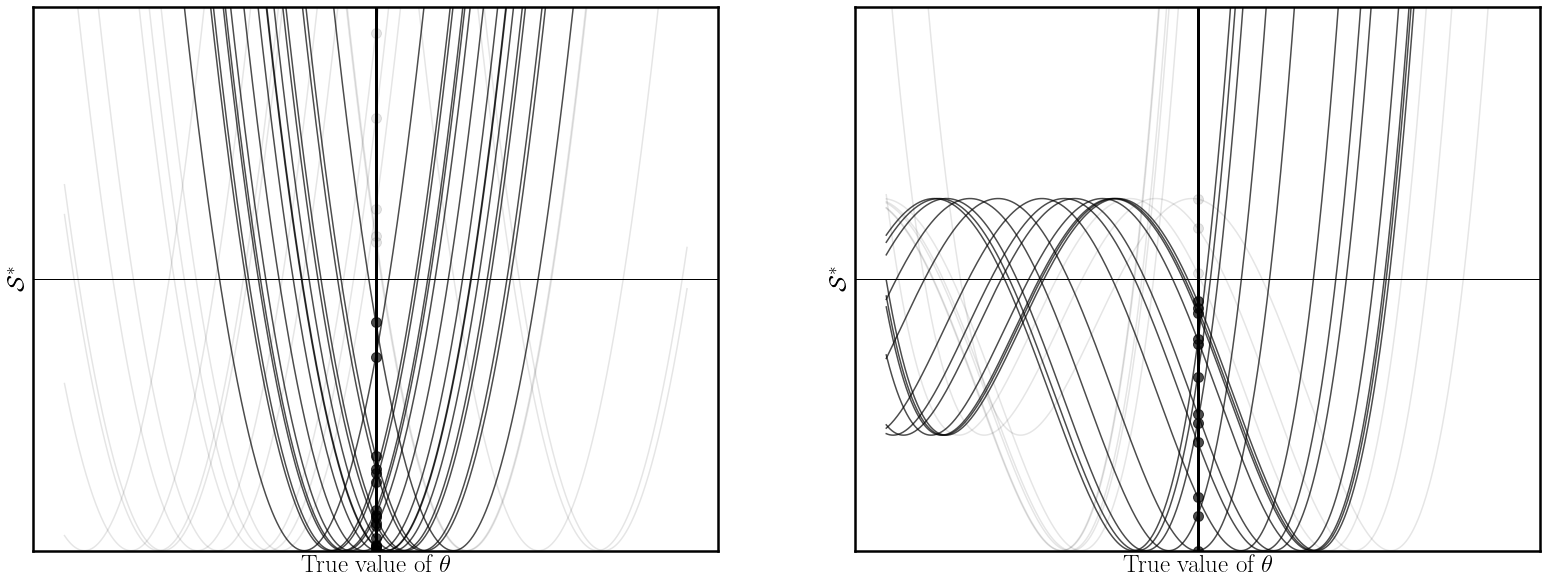}
    \caption{Evolution of the statistic $\mathcal{S}$ as a function of the model parameter $\theta$ from different pseudo-experiments with fixed true value $\bar{\theta}$. Vertical line shows the true value of $\theta$, while the horizontal one shows the threshold $\mathcal{S}^*$ for the statistic such that $\mathcal{S}(\bar{\theta}) \leq \mathcal{S}^* $ only $x\%$ of the time. Black curves are those that fulfill this condition while grey ones are those that do not. In~the left plot the intersection between the curves and the line $\mathcal{S} = \mathcal{S}^*$ defines CIs which by construction \textit{cover} the true value of $\theta$ $x\%$ of the time. These CIs cannot be anymore constructed for the curves in the right plot where the statistic has below $ \mathcal{S}^*$  more than one minimum. In~both plots the curves shown are not specific on any particular problem but only serve as a schematic~representation.  }
    \label{fig:Statistc_cov}
\end{figure}

In the frequentist approach, these bounds are obtained by looking at the log-run behavior of the statistic: a threshold value $\mathcal{S}^*$ of the statistic is defined such that in infinite experiments with fixed parameter $\bar{\theta}$, we would have observed $\mathcal{S} \leq \mathcal{S}^* $ only $x\%$ of the time. The~lower or upper bound $\theta_x$ is then defined such that $\mathcal{S}(\theta_x) =  \mathcal{S}^*$. In~other words, we look for the values of the parameter that are excluded with a $x\%$ CL.  The~statistic $\mathcal{S}$ is generally constructed to increase monotonically for values of $\theta$ smaller or bigger than the best estimated value $\hat{\theta}$, which is by definition the value whose exclusion can be claimed with a $0\%$ CL. For~an UL (LL) this means that values of $\theta$ bigger (smaller) than $\theta_x$ are excluded with higher CL and they are therefore less likely to be true. This is schematically shown in the left panel of Figure~\ref{fig:Statistc_cov} where the statistic $\mathcal{S}$ is shown as a function of the parameter $\theta$ for different experiments in which $\theta$ is fixed to the true value $\bar{\theta}$ (vertical line). 

By searching for $\theta_x$ such that   $\mathcal{S}(\theta_x) =  \mathcal{S}^*$ we obtain a LL $\theta_x^{LL}$ and UL $\theta_x^{UL}$. By~construction only $x \%$ of these curves have $\mathcal{S}(\bar{\theta}) \leq  \mathcal{S}^*$, which are shown in black in the left plot of Figure~\ref{fig:Statistc_cov}, while the remaining curves are shown in grey. This implies that the true value $\bar{\theta}$ lies in the interval $\left[ \theta_x^{LL}, \theta_x^{UL}  \right]$ $x\%$ of the time. Such interval is referred to as confidence interval (CI) and it is said to \textit{cover} the true value of $\theta$ $x\%$ of the~time.
The condition  that values more extreme than the obtained bounds  are rejected with more CL applies for the  analysis described in Sections~\ref{Subsec:b0}--\ref{Subsec:b2}, but~in general it is not always true, as~shown schematically in the right plot of Figure~\ref{fig:Statistc_cov}. In~such cases LLs, ULs and CIs do not have a straightforward interpretation. This is the reason why it is good practice to report in the result of the statistical analysis also the likelihood shape as a function of the free~parameters.

If $x$ is chosen to be $68$, the~interval between the lower and upper bounds defines the so-called $68\%$ CI. When the background is estimated from an OFF measurement (see Section~\ref{Subsec:b2}) we can use the statistic  defined in Equation~(\ref{Eq:Lklratio_b2}) which is a $\chi^2$ random variable with 1 degree of freedom\endnote{The CDF of a $\chi^2$ distribution with 1 degree of freedom is 0.68 for $\chi^2 =  1$ and 0.95 for $\chi^2 =  3.84$.}. By~looking for the bounds for which $\mathcal{S} =  1$ we obtain  the $68\%$ CI  of $s$,  that for large count numbers is given by
\begin{equation}
    \left[  \hat{s} - \sqrt{ N_{on} + \alpha^2 N_{off}} \; , \; \hat{s} + \sqrt{ N_{on} + \alpha^2N_{off}} \right]
\end{equation}
where $\hat{s}$ is the estimated signal given by $N_{on} - \alpha N_{off}$.

\begin{figure}[H]
    
    \includegraphics[width=0.97\linewidth]{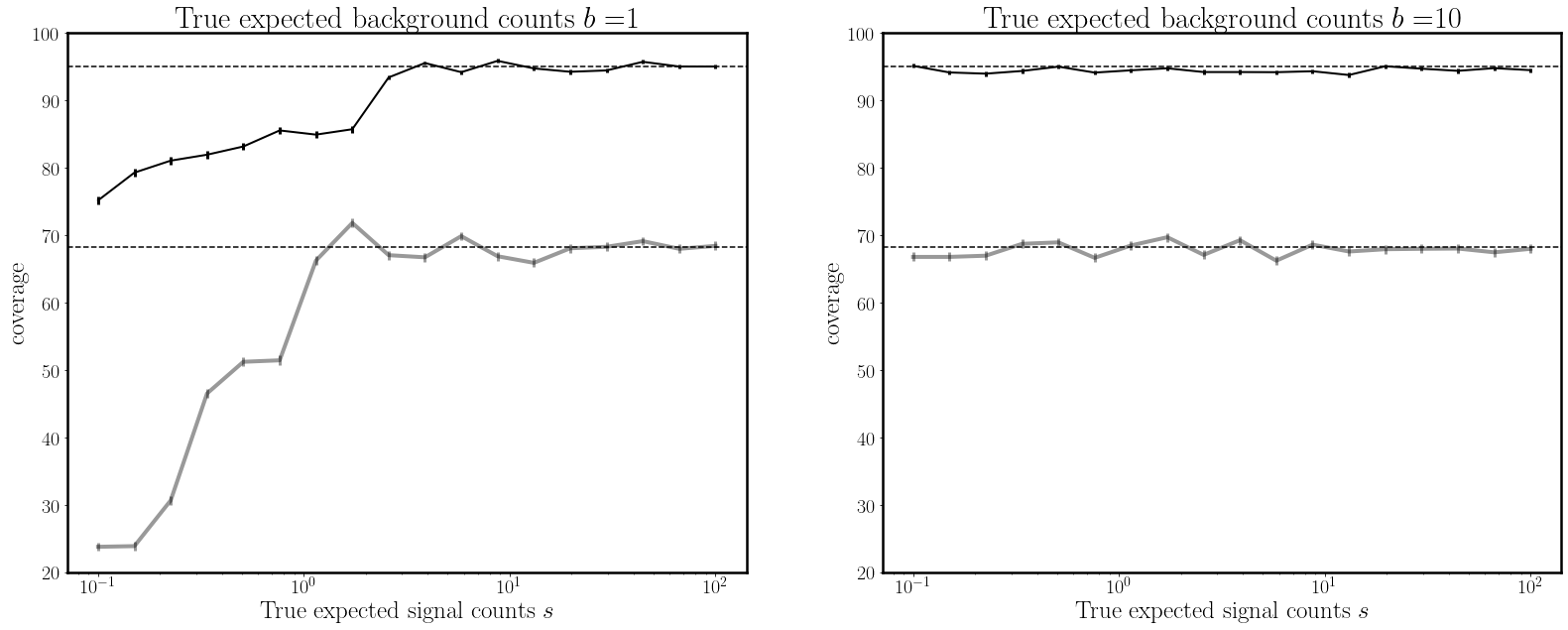}
    \caption{Evolution of the coverage of the CIs obtained from solving $\mathcal{S} =$ 1 (grey line) or $\mathcal{S} =$ 3.84 (black line) as a function of the signal $s$.  $\mathcal{S}$ is the statistic defined in  Equation~(\ref{Eq:Lklratio_b2}). The~dashed horizontal lines show the expected coverage from the assumption that $\mathcal{S} $ is a $\chi^2$-random variable with 1 degree of freedom. In~each MC simulation the observed counts $N_{on}$ and $N_{off}$ were simulated from a Poisson distribution with expected counts $s+\alpha b$ and $b$, respectively. \textbf{Left panel}: the expected background count is fixed to 1. \textbf{Right panel}: the expected background count is fixed to 10.   In~both plots $\alpha = 0.5$ is~assumed. }
    \label{fig:coverage}
\end{figure}

When looking for the UL, the~CL is usually set to $95\%$, with~other common values being $90\%$ or $99.9\%$. In~this case the UL $s_{95}$ is obtained by solving $\mathcal{S}(s_{95}) =  3.84$. The~coverage of the $68\%$ and $95\%$ CI is shown in Figure~\ref{fig:coverage} for different true signal and background counts. As~one can see from this figure by imposing $\mathcal{S}=$ 1 or 3.84 we have a good coverage (of $68\%$ and $95\%$, respectively) for large enough counts. Although~when the counts are too small the CIs tend to \textit{undercover} the true value of $s$. Such problem is well-known and it requires ad hoc adjustments~\cite{rolke2001confidence,rolke2005limits} in order to recover the desired coverage. 

In the Bayesian context, the concept of \textit{coverage} is meaningless, since the objective of the analysis is not to look for the long-run performance of a given statistic, but~to provide a probabilistic statement on the model parameters. In~this case, CIs are replaced by credible intervals whose purpose is to provide the analyzer an interval where the model parameter lies with a given probability. Let us assume that we are interested in finding the $68\%$ credible interval $[s_1, s_2]$ of the signal $s$. By~using the PDF of $s$ in Equation~(\ref{Eq:Integral_b}) such interval is defined as follows
\begin{equation}
 \int_{s_1}^{s_2}   p( s  \; | \;  N_{on},  N_{off}; \alpha) \; ds  = 0.68 
 \label{Eq:Credible}
\end{equation}
where $s_1$ and $s_2$ can be chosen\endnote{Another way to choose  $s_1$ and $s_2$ is to guarantee that the mean is the central value of the interval $[s_1, s_2]$. In~principle, one is free to pick up infinitely intervals from the constraint given by Equation~(\ref{Eq:Credible}). A~more detailed discussion on this topic can be found in Ref.~\cite{o2004kendall}. } to include the values of highest probability density. Similarly the $95\%$ UL $s_{UL}$ on $s$ is obtained from
\begin{equation}
 \int_{s_{UL}}^{\infty}   p( s  \; | \;  N_{on},  N_{off}; \alpha) \; ds  = 0.05.
 \label{Eq:Credible2}
\end{equation}

A comparison between the confidence and credible intervals, computed with the frequentist and Bayesian approach, respectively, can be found in Refs.~\cite{knoetig2014signal, casadei2014objective, d2021signal}. When comparing them it is although important to remember that the two approaches are providing the answer to two completely different questions. In~the frequentist case the analyzer is given a procedure for computing the interval that in infinite experiments will cover the true value of the parameters a desired fraction of the time. The~parameter is fixed in these infinite experiments and the coverage is usually checked by performing MC simulations. In~the Bayesian approach instead the model parameters are not fixed and they lie in the computed  interval  with a given~probability.  

%


\section{Flux Estimation and Model Parameter~Inference}
\label{Sec:Flux}

We have reached the final step of the inference analysis (see Figure~\ref{fig:Scheme}) which started in Section~\ref{Sec:EventRecon} from the shower image data: estimate the source flux and model parameters. Starting point of this analysis is the expected signal count $s$, whose estimation from the events list is described in Section~\ref{Sec:detection}. Taking into account the exposure of the observation given by the energetic ($E$), temporal ($t$) and solid angle ($\Omega$) range (hereafter denote by $\Delta$) in which the events have been collected we have
\begin{equation}
    s = \int_{\Delta} \Phi' (E_r, \hat{\mathbf{n}}_r, t) d E_r d\hat{\mathbf{n}}_r dt 
    \label{Eq:source_exp_int}
\end{equation}
where  $\Phi'$ is the differential observed flux with units of 1/( solid angle $\times$ time $\times$ energy), while $E_r$ and $\hat{\mathbf{n}}_r$ are the reconstructed energy and arrival direction (for the time a perfect temporal resolution is assumed being of the order of hundreds of nanoseconds). The~observed flux is given by the convolution of the differential source flux $\Phi$ with the IRF of the~telescope:
\begin{equation}
    \Phi' (E_r, \hat{\mathbf{n}}_r, t) = \int_{E} \int_{\Omega} \Phi (E, \hat{\mathbf{n}}, t) \cdot \text{IRF}(E_r, \hat{\mathbf{n}}_r  ,   E, \hat{\mathbf{n}} ) \; d E \; d\hat{\mathbf{n}}.
    \label{Eq:PhiObse_1}
\end{equation}

The IRF can be though as the probability of detecting  a photon with energy $E$ and arrival direction $\hat{\mathbf{n}}$ and assigning to it a reconstructed energy and arrival direction $E_r$ and $\hat{\mathbf{n}}_r$, respectively.  Following the rules of conditional probability the IRF can be expanded as~follows:
\begin{equation}
\text{IRF}(E_r, \hat{\mathbf{n}}_r  ,   E, \hat{\mathbf{n}} ) = \text{IRF}(E_r \; | \; \hat{\mathbf{n}}_r  ,   E, \hat{\mathbf{n}} ) \cdot \text{IRF}(\hat{\mathbf{n}}_r \; | \;     E, \hat{\mathbf{n}} ) \cdot \text{IRF}(  E, \hat{\mathbf{n}}). 
\end{equation}

Since $E_r$ and $\hat{\mathbf{n}}_r$ are conditional independent variables\endnote{Generally the performance of the energy and direction reconstruction only depends on the event true energy and arrival direction, which justifies  the assumption that $E_r$ and $\hat{\mathbf{n}}_r$ are conditional independent variables.} given $E$ and $\hat{\mathbf{n}}$, the~above expression can be rewritten as 
\begin{equation}
    \text{IRF}(E_r, \hat{\mathbf{n}}_r  ,   E, \hat{\mathbf{n}} ) = D( E_r \; | \;  E, \hat{\mathbf{n}})  \cdot \text{PSF}( \hat{\mathbf{n}}_r \; | \;  E, \hat{\mathbf{n}})  \cdot \varepsilon( E, \hat{\mathbf{n}}),
\end{equation}
where we have identified the first term with the energy dispersion $D$, the~second one with the point spread function (PSF) and the last term with the collection efficiency  $\varepsilon$ of the telescopes. To date, we have made the assumption that the observation parameters (such as the zenith angle of the observation) are (or can be assumed to be) constant during the~observation.

We can further simplify the IRF expression by ignoring (i.e., by~integrating out) the information on the arrival direction $\hat{\mathbf{n}}$, thus reducing the dimension of the problem from 3 to 1. This assumption is justified if, for~instance, the~observation is performed on a point-like source, which will be also assumed hereafter to be steady. 
Having simplified our problem\endnote{For a more accurate discussion that includes  also other variables (such as the photon direction $\hat{\mathbf{n}}$) one can check Ref.~\cite{ackermann2012fermi}.}, Equations~(\ref{Eq:source_exp_int}) and (\ref{Eq:PhiObse_1})  can be then rewritten, respectively, as
\begin{align}
s &= \int_{\Delta} \Phi' (E_r) d E_r \label{Eq:s1} \\ 
    \Phi' (E_r ) &= \int_{E}  \Phi (E ) \cdot  D( E_r \; | \;  E ) \cdot \varepsilon( E )  d E. 
\label{Eq:s2}
\end{align}

In order to get the flux $\Phi$ from the expected counts $s$ two approaches are used: \textit{unfolding} and \textit{forward folding}.
\subsection{Unfolding}
\label{Sec:Unfolding}
If we divide the energy range in bins, the~expected counts of gamma ray in the $i$-th bin $\Delta_i$ of reconstructed energy is (when combining Equations~(\ref{Eq:s1}) and (\ref{Eq:s2}))
\begin{equation}
   s_i = \int_{\Delta_i} d E_r \sum_{j} \int_{\Delta_j} d E \Phi(E) D(E_r | E) \varepsilon(E) \equiv \sum_{j} R_{i,j} \bar{s}_{j},
   \label{Eq:_si}
\end{equation}
where $\bar{s}_{j}$ is the expected number of gamma rays from the source flux in the $j$-th bin $\Delta_j$, i.e.,
\begin{equation}
    \bar{s}_{j} = \int_{\Delta_j}  \Phi(E) d E.
\end{equation}

The matrix $R_{i,j}$ is the \textit{response matrix} which is the probability of detecting (due to the collection efficiency $\varepsilon$) a photon with energy in the range $\Delta_j$ and assign to it (due to the energy dispersion $D$) a different energy bin $\Delta_i$. 
From Equation~(\ref{Eq:_si}) its expression in formula is given by
\begin{equation}
    R_{i,j} = \frac{ 1}{ \bar{s}_j} \int_{\Delta_i} d E_r \int_{\Delta_j} d E \Phi(E) D(E_r | E) \varepsilon(E)  ,
    \label{Eq:response}
\end{equation}
while in practice $R_{i,j}$ is estimated from MC simulations as
\begin{equation}
    R_{i,j} = \frac{n_{i,j}^{\gamma}}{N_j^{\gamma}},
\end{equation}
where $N_j^{\gamma} $ is the total number of gamma ray events simulated (according to an assumed source flux $\Phi$) with true energy in the energy range $\Delta_j$, and~$n_{i,j}^{\gamma}$ the number of those same events that have been detected by the telescope with a reconstructed energy in the energy range $\Delta_i$. 
Clearly, by~summing over $i$ we recover the binned collection efficiency
\begin{equation}
    \varepsilon_j = \sum_i R_{i,j}.
\end{equation}

\begin{figure}[H]
    
    \includegraphics[width=0.45\linewidth,height=0.43\linewidth]{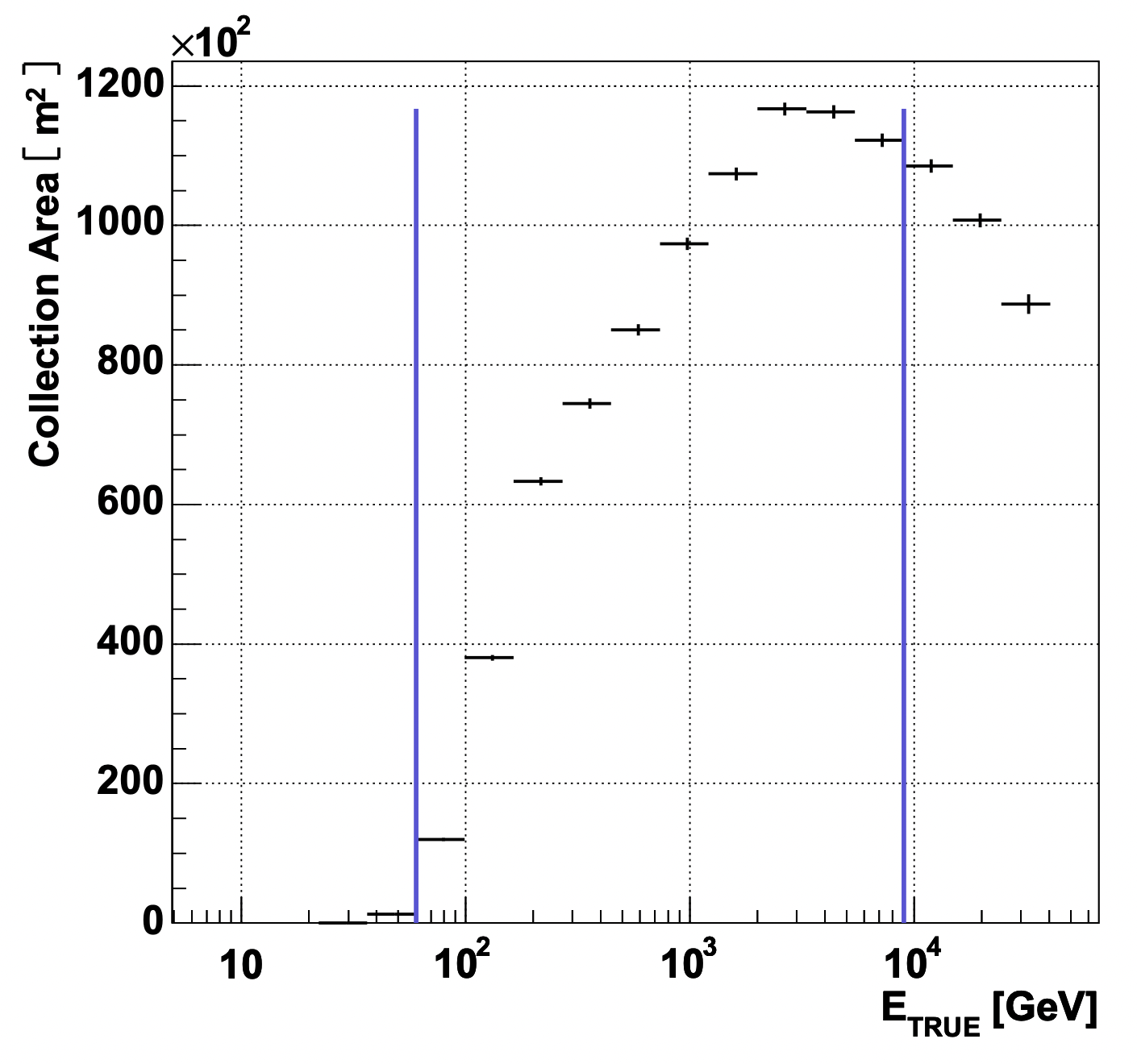}
        \includegraphics[width=0.45\linewidth,height=0.43\linewidth]{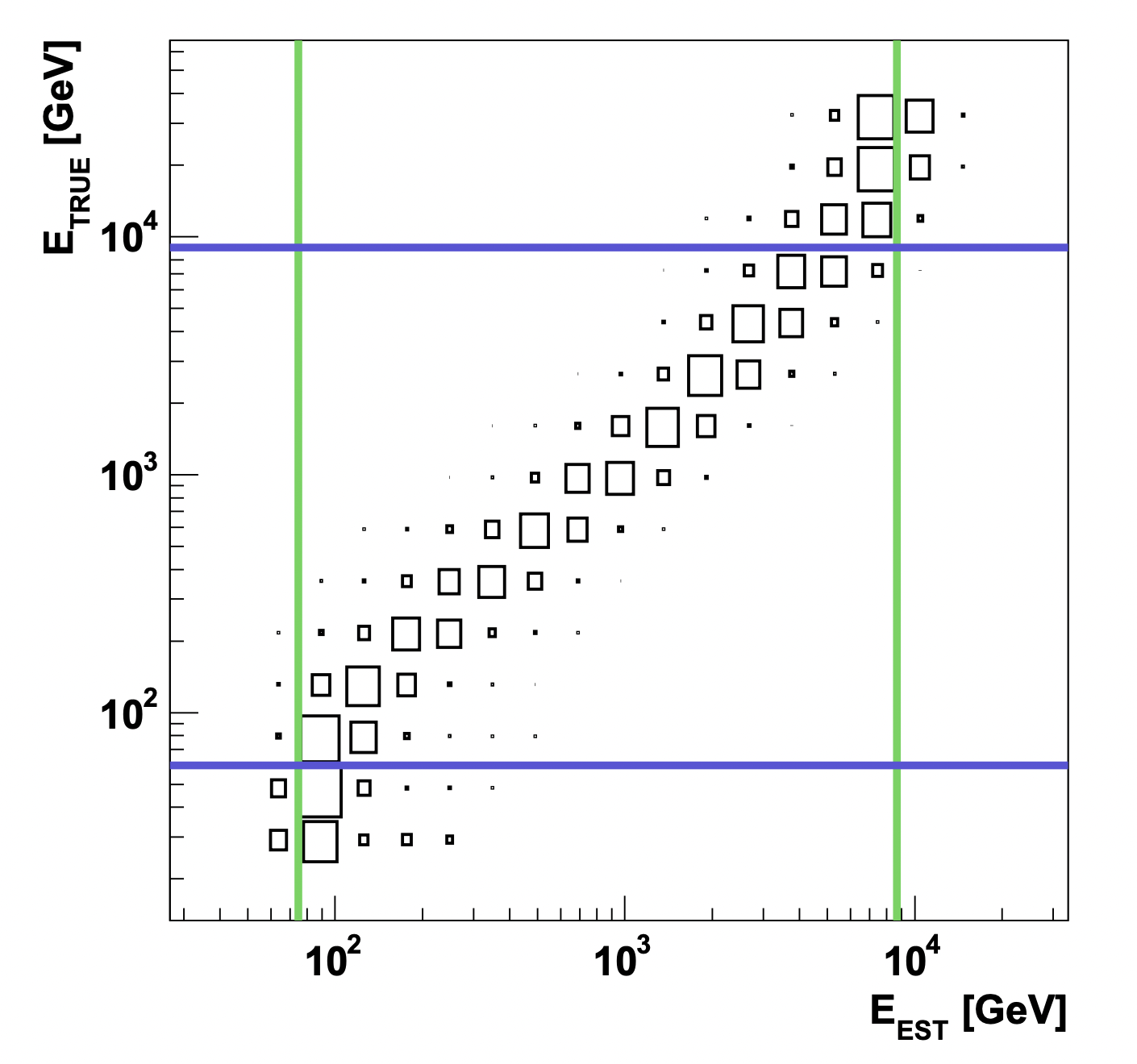}
    \caption{\textbf{Left panel}: Evolution in energy of the collection efficiency $\varepsilon(E)$ multiplied by the collection area of the telescope, which for IACTs is generally of the order of $10^5\,\text{m}^2$.  \textbf{Right panel}: evolution in reconstructed/estimated energy and true energy of the binned dispersion energy (or migration matrix).  Both figures are Reprinted with permission from Ref.~\cite{albert2007unfolding} Copyright 2007 Albert et al.  }
    \label{fig:CollectionDisp}
\end{figure}

An example of the binned collection efficiency along with the energy dispersion from the same experiment can be found in Figure~\ref{fig:CollectionDisp}.

Goal of the unfolding procedure is to find a solution of Equation~(\ref{Eq:_si}), by~inverting the response matrix
\begin{equation}
    \bar{s}_j = \sum_{i} R^{-1}_{j,i} s_i.
    \label{Eq:unfol_matrix}
\end{equation}

Thus, unfolding is basically a deconvolution problem and shares its typical problems, like the fact that the response matrix is, in~general, non-invertible. 
As all ill-posed problems \textit{regularization} procedures are required in order to find a solution and to prevent overfitting. In~the context of IACTs analysis, common regularization procedures are those of  Tikhonov~\cite{tikhonov1963solution}, Bertero~\cite{bertero1989linear} and Schmelling~\cite{schmelling1994method}. For~a more detailed discussion and comparison of these approaches with applications to data collected with the MAGIC telescopes see Ref.~\cite{albert2007unfolding}. It is good practice to show the unfolding result with several of these approaches to cross-check the reconstructed flux and to also report along with the reconstructed flux points $\bar{s}_j$ their correlation matrix. Such a correlation matrix is needed if one is willing to fit the flux points $\bar{s}_j$ with a spectral~model. 

To date, the unfolding approach has been discussed as a geometrical problem: given a known vector $s$ and a known matrix $R$, we wish to invert $R$ in order to find the unknown vector $\bar{s}$ as shown in Equation~(\ref{Eq:unfol_matrix}). In~the \textit{Bayesian unfolding} approach instead the problem is a probabilistic one: given our prior knowledge $I$ and the expected counts $s_i$ in the reconstructed energy bins, we wish to get the probability distribution of $\bar{s}_j$
\begin{equation}
    p( \bar{s}_j \, |\, s_i , I ) = \frac{ p(  s_i \, |\,\bar{s}_j , I) \,  \cdot \, p(\bar{s}_j |  I )}{\sum_i p(  s_i \, |\,\bar{s}_j , I) \, \cdot \, p(\bar{s}_j |  I ) }.
    \label{Eq:Bayes_unfolding}
\end{equation}

The \textit{prior} $p(\bar{s}_j |  I )  $ is the binned normalized ( $\sum_j p(\bar{s}_j |  I ) = 1  $) flux that we initially assumed for the source, while the term  $p(  s_i \, |\,\bar{s}_j , I)$ is the probability of measuring an expected signal count in the reconstructed energy bin $\Delta_i$ given the \textit{true} signal count $\bar{s}_j$ in the energy bin $\Delta_j$. This term is related to the response matrix defined in Equation~(\ref{Eq:response}). 

An \textit{iterative} method for getting the \textit{posterior} in Equation~(\ref{Eq:Bayes_unfolding}) that takes as a prior $p(\bar{s}_j |  I )  $  the posterior obtained from a previous iteration can be found in Ref.~\cite{d1995multidimensional} and later on revised and improved by the same author in Ref.~\cite{d2010improved} . More recently the author of Ref.~\cite{choudalakis2012fully} proposed a \textit{fully Bayesian unfolding} with applications to numerous examples. 

\subsection{Forward~Folding}
The main advantage of the unfolding algorithm is its ability to show a distribution  that is as much as possible equivalent to the observed distribution of events with physical and instrumental effects being removed.  Although~some assumptions about the flux are inevitable as discussed in  Section~\ref{Sec:Unfolding}, the~desired outcome of the unfolding procedure is to ``interpret'' the data as little as~possible.

In the total opposite direction, we can find the forward folding approach. In~this case, a parametric model for the intrinsic flux is assumed and the final result is to provide an estimate of the free model parameters $\theta$. 
When it is reasonable to believe that the source flux can be described by one or a family of parametric functions, the~forward folding is always preferable to the unfolding one, being easier to implement and free of those problems that are typical of the unfolding methods which are cured by regularization. The~problem falls therefore within the realm of ``fitting'' problems: searching for the values $\hat{\theta}$ that maximize the likelihood function, defined as the probability of getting the observed data given the model parameters $\theta$. The~observed data are the list of $N_{on}$ events (obtained from the shower image as discussed in Section~\ref{Sec:EventRecon}) with their reconstructed energy, arrival direction and time (hereafter denoted for simplicity $x$). If~the background is estimated from an OFF region (see Section~\ref{Subsec:b2}) one has to take into account also the observed counts $N_{off}$ in the OFF region and the normalization factor $\alpha$. The~likelihood function is
\begin{align}
        & \mathcal{L}(\theta \, | \pi, \, N_{on}, N_{off}, \alpha, x_1, \dots, x_{N_{on}})  = \nonumber \\
        &  p(\pi \, | \, \theta) \; \cdot \; \mathcal{P}( N_{on} | s + \alpha b ) \cdot \mathcal{P}( N_{off} | b ) \cdot \prod_{i =1}^{N_{on}} \:  \left(  \frac{ f_s(x_i \,| \,  \theta, \pi )  +  f_b(x_i ) }{s + \alpha b} \right) 
        \label{Eq:Unbinned_Lkl}
\end{align}
where $\pi$ are the nuisance parameters of the model and $p(\pi \, | \, \theta)  $ their probability distribution given $\theta$. The~function $f_s$ is the differential source flux with the IRF of the telescopes being taken into account, such that
\begin{equation}
    \int dx \; f_s(x \,| \,  \theta, \pi ) = s,
    \label{Eq:Integral_of_s}
\end{equation}
where $s$ is the expected signal counts in the ON region. The~expected background count $b$ in the OFF region is instead provided by
\begin{equation}
    \int dx \; f_b(x ) = b
\end{equation}
where $f_b$ is the differential background template model. The~function $\mathcal{P}$ is the Poisson distribution defined in Equation~(\ref{Eq:Poisson}). The~likelihood function defined in Equation~(\ref{Eq:Unbinned_Lkl}) is usually referred to as ``unbinned likelihood'' to distinguish it from its binned version
\begin{align}
   \mathcal{L}(\theta \, | \pi, \, N_{on}^{(1)},\dots, N_{off}^{(1)}, \dots, \alpha)  =
 p(\pi \, | \, \theta) \; \cdot  \prod_{i =1}^{\text{all bins}}  \mathcal{L}_i(  \theta \, | \pi, \, N_{on}^{(i)}, N_{off}^{(i)}, \alpha),
        \label{Eq:Binned_Lkl}
\end{align}
where $\mathcal{L}_i $ is the likelihood of the single $i$-th bin (in which  $ N_{on}^{(i)}$ and  $ N_{off}^{(i)}$ counts have been observed in the ON and OFF region, respectively) given by
\begin{equation}
 \mathcal{L}_i(  \theta \, | \pi, \, N_{on}^{(i)}, N_{off}^{(i)} , \alpha) =  \mathcal{P}( N_{on}^{(i)} \; |\,  s_i(\theta, \pi) + \alpha b_i ) \cdot \mathcal{P}( N_{off}^{(i)}  \; |\, b_i ).
\end{equation}

The variable $b_i$ has to be treated as a nuisance parameter and fixed to the value given in Equation~(\ref{Eq:hathatb}) if the frequentist approach is implemented or integrated out in the Bayesian one. The~variable $s_i(\theta, \pi) $ is instead obtained by the integral in Equation~(\ref{Eq:Integral_of_s}) with integration limits on $x$ defined by the bin under~consideration. 

The forward folding procedure applied in IACTs can be found for instance in fundamental physics studies, such as the search for dark matter~\cite{rico2020gamma} or tests of the equivalence principle from the time of flight of cosmic gamma rays~\cite{acciari2020bounds}. 

\section{Discussion}

From the discovery of the TeV emission from the Crab nebula in 1989 by the  Whipple collaboration~\cite{weekes1989observation}, IACTs have been able in recent decades to give birth to a mature field of gamma ray astronomy. Instruments such as MAGIC, HESS, and VERITAS  discovered numerous  astrophysical sources at TeV energies, allowing investigation of the physics of remote cosmic objects.  Apart from the technological development needed for the construction and maintenance of these telescopes, a~huge effort has been carried  out in recent decades to adapt and explore statistical tools aimed at extracting all of the information contained in the collected~data.

The most challenging issue in the analysis of IACTs data is the predominant presence of background events  that require detailed studies such as the estimation of the background from OFF regions as discussed in Section~\ref{Subsec:b2}. Gamma rays only compose a small fraction of the cosmic flux that hits our atmosphere producing the Cherenkov light observed by the telescopes. Techniques such as the ``multivariate analysis'' (see Section~\ref{Sec:Multivar}) or the CNNs (see Section~\ref{Sec:CNNs}  ) are the current \textit{state of the art} for discriminating gamma rays events from the~background. 

Another important factor that makes the statistical analysis so important and challenging for these instruments is that unfortunately we do not have a pure source of gamma rays, which is steady and under our control, and~can then be used for calibrating the telescopes. The~closest we have to a steady and bright gamma ray source is the Crab Nebula~\cite{amato2021crab}, which is indeed used as a standard for calibrating the instrument whenever an IACT observatory starts its operations. In~order to infer instrumental properties, such as the energy threshold, a~signal from the Crab Nebula is collected and then compared with the expected (obtained from MC simulations) response. MC simulations are therefore of huge importance for IACTs and, moreover, they also provide a way for training the BTD and RF algorithms briefly discussed in Section~\ref{Sec:Multivar}. The~problem is that instrumental effects (such as the  reflectance of the mirrors) and the atmospheric conditions have to be taken into account in these MC simulations, which in most cases requires  approximations and idealized instrumental parametrizations. This is the reason why different efforts were made  as discussed in Section~\ref{Sec:MC} for making these MC simulations as realistic as~possible. 

Once the above issues are overcome, we have to quantify, given the list of  detected events, how likely it is that a flux of gamma rays has been detected and how confidently we can set some values to such a flux. This has been discussed in Section~\ref{Sec:detection} where we showed, using the frequentist and Bayesian approach, different solutions to this problem, emphasizing with examples their differences and mentioning some of the most recent developments.   Lastly in Section~\ref{Sec:Flux}, the IRF is taken into account in order to provide a flux estimation which is as much as possible similar to the intrinsic flux of gamma~rays. 

With the construction of CTA~\cite{doi:10.1142/10986}, the~next generation of IACTs has ten times more sensitivity than the current instruments, and the~statistical tools described in this review will become more and more indispensable in order to put the capability of the telescopes at its~limits.


\vspace{6pt} 


\funding{This review work was funded by  the Research Council of Norway grant, project number~301718.}

\institutionalreview{Not applicable.}

\informedconsent{Not applicable.}

\dataavailability{Not applicable.} 

\acknowledgments{ We would like to thank 
Julia Djuvsland 
 for useful comments on a early version of this manuscript and the anonymous reviewers for their insightful inputs and~suggestions.}

\conflictsofinterest{The author declares no conflict of~interest.} 



\abbreviations{Abbreviations}{
The following abbreviations are used in this manuscript:\\

\noindent 
\begin{tabular}{@{}ll}
BDT & Boosted Decision Tree \\ 
BF & Bayes Factor \\
CDF  & Cumulative Distribution Function \\
CI & Confidence Interval \\
CL & Confidence Level\\
CNN & Convolutional Neural Networks\\
DL & Deep Learning \\
IACT & Imaging Atmospheric Cherenkov Telecope\\
ImPACT & Image Pixel-wise Atmospheric Cherenkov Telescopes\\
IRF & Instrument Response Function \\
LL & Lower Limit \\
MC & Monte Carlo\\
PDF   & Probability Distribution Function \\
PhE & Photo-Electron \\
PMF   & Probability Mass Function \\
PSF  & Point Spread Function \\
RF & Random Forest \\
ROI & Region of Interest \\
UL & Upper Limit \\
\end{tabular}}

\appendixtitles{no} 
\begin{adjustwidth}{-\extralength}{0cm}
\printendnotes[custom]
%
%

\reftitle{References}

%

\end{adjustwidth}

\begin{thebibliography}{999}

\bibitem[Aleksi{\'c} \em{et~al.}(2016)Aleksi{\'c}, Ansoldi, Antonelli,
Antoranz, Babic, Bangale, Barcel{\'o}, Barrio, Gonz{\'a}lez, Bednarek,
et~al.]{aleksic2016major}
Aleksi{\'c}, J.; Ansoldi, S.; Antonelli, L.A.; Antoranz, P.; Babic, A.;
Bangale, P.; Barcel{\'o}, M.; Barrio, J.; Gonz{\'a}lez, J.B.; Bednarek, W.;
et~al.
\newblock The major upgrade of the MAGIC telescopes, Part II: A performance
study using observations of the Crab Nebula.
\newblock {\em Astropart. Phys.} {\bf 2016}, {\em 72},~76--94. [\href{http://doi.org/10.1016/j.astropartphys.2015.02.005}{CrossRef}]

\bibitem[Hinton(2004)]{hinton2004status}
Hinton, J.
\newblock The status of the HESS project.
\newblock {\em New Astron. Rev.} {\bf 2004}, {\em 48},~331--337. [\href{http://dx.doi.org/10.1016/j.newar.2003.12.004}{CrossRef}]


\bibitem[Holder \em{et~al.}(2006)Holder, Atkins, Badran, Blaylock, Bradbury,
Buckley, Byrum, Carter-Lewis, Celik, Chow, et~al.]{holder2006first}
Holder, J.; Atkins, R.; Badran, H.; Blaylock, G.; Bradbury, S.; Buckley, J.;
Byrum, K.; Carter-Lewis, D.; Celik, O.; Chow, Y.;  et~al.
\newblock The first VERITAS telescope.
\newblock {\em Astropart. Phys.} {\bf 2006}, {\em 25},~391--401. [\href{http://dx.doi.org/10.1016/j.astropartphys.2006.04.002}{CrossRef}]

\bibitem[Actis \em{et~al.}(2011)Actis, Agnetta, Aharonian, Akhperjanian,
Aleksi{\'c}, Aliu, Allan, Allekotte, Antico, Antonelli,
et~al.]{actis2011design}
Actis, M.; Agnetta, G.; Aharonian, F.; Akhperjanian, A.; Aleksi{\'c}, J.; Aliu,
E.; Allan, D.; Allekotte, I.; Antico, F.; Antonelli, L.;  et~al.
\newblock Design concepts for the Cherenkov Telescope Array CTA: An advanced
facility for ground-based high-energy gamma-ray astronomy.
\newblock {\em Exp. Astron.} {\bf 2011}, {\em 32},~193--316. [\href{http://dx.doi.org/10.1007/s10686-011-9247-0}{CrossRef}]

\bibitem[Wasserstein and Lazar(2016)]{doi:10.1080/00031305.2016.1154108}
Wasserstein, R.L.; Lazar, N.A.
\newblock The ASA Statement on p-Values: Context, Process, and Purpose.
\newblock {\em  Am. Stat.} {\bf 2016}, {\em 70},~129--133. [\href{http://dx.doi.org/10.1080/00031305.2016.1154108}{CrossRef}]

\bibitem[Neyman and Pearson(1933)]{neyman1933ix}
Neyman, J.; Pearson, E.S.
\newblock IX. On the problem of the most efficient tests of statistical
hypotheses.
\newblock {\em Philos. Trans. R. Soc. Lond. Ser. A  Contain. Pap. A Math. Phys. Character} {\bf
1933}, {\em 231},~289--337.

\bibitem[Wilks(1938)]{Wilks}
Wilks, S.
\newblock {The Large-Sample Distribution of the Likelihood Ratio for Testing
Composite Hypotheses}.
\newblock {\em Ann. Math. Stat.} {\bf 1938}, {\em 9},~60--62. [\href{http://dx.doi.org/10.1214/aoms/1177732360}{CrossRef}]

\bibitem[Protassov \em{et~al.}(2002)Protassov, Van~Dyk, Connors, Kashyap, and
Siemiginowska]{protassov2002statistics}
Protassov, R.; Van~Dyk, D.A.; Connors, A.; Kashyap, V.L.; Siemiginowska, A.
\newblock Statistics, handle with care: Detecting multiple model components
with the likelihood ratio test.
\newblock {\em  Astrophys. J.} {\bf 2002}, {\em 571},~545. [\href{http://dx.doi.org/10.1086/339856}{CrossRef}]

\bibitem[Loredo(1990)]{loredo1990laplace}
Loredo, T.J.
\newblock From Laplace to supernova SN 1987A: Bayesian inference in
astrophysics. In {\em Maximum Entropy and Bayesian Methods}; Springer: Berlin/Heidelberg, Germany,  1990;
pp. 81--142.

\bibitem[Lyons(2008)]{lyons2008open}
Lyons, L.
\newblock Open statistical issues in particle physics.
\newblock {\em  Ann. Appl. Stat.} {\bf 2008}, {\em 2},~887--915. [\href{http://dx.doi.org/10.1214/08-AOAS163}{CrossRef}]

\bibitem[van~de Schoot \em{et~al.}(2021)van~de Schoot, Depaoli, King, Kramer,
M{\"a}rtens, Tadesse, Vannucci, Gelman, Veen, Willemsen,
et~al.]{van2021bayesian}
van~de Schoot, R.; Depaoli, S.; King, R.; Kramer, B.; M{\"a}rtens, K.; Tadesse,
M.G.; Vannucci, M.; Gelman, A.; Veen, D.; Willemsen, J.;  et~al.
\newblock Bayesian statistics and modelling.
\newblock {\em Nat. Rev. Methods Prim.} {\bf 2021}, {\em 1},~1--26. [\href{http://dx.doi.org/10.1038/s43586-020-00001-2}{CrossRef}]

\bibitem[V{\"o}lk and Bernl{\"o}hr(2009)]{cite-key}
V{\"o}lk, H.J.; Bernl{\"o}hr, K.
\newblock Imaging very high energy gamma-ray telescopes.
\newblock {\em Exp. Astron.} {\bf 2009}, {\em 25},~173--191.\linebreak s10686-009-9151-z. [\href{http://dx.doi.org/10.1007/s10686-009-9151-z}{CrossRef}]

\bibitem[{Hillas}(1985)]{1985ICRC....3..445H}
{Hillas}, A.M.
\newblock {Cerenkov Light Images of EAS Produced by Primary Gamma Rays and by
Nuclei}.
\newblock In~{Proceedings of the} 19th International Cosmic Ray Conference (ICRC19), San Diego, CA, USA, 11--23 August 1985;
Volume 3,    p. 445.

\bibitem[Zanin \em{et~al.}(2013)Zanin, Carmona, Sitarek, Colin, Frantzen, Gaug,
Lombardi, Lopez, Moralejo, Satalecka, et~al.]{zanin2013mars}
Zanin, R.; Carmona, E.; Sitarek, J.; Colin, P.; Frantzen, K.; Gaug, M.;
Lombardi, S.; Lopez, M.; Moralejo, A.; Satalecka, K.;  et~al.
\newblock MARS, the MAGIC analysis and reconstruction software.
\newblock   In~{Proceedings of the} 33st International Cosmic Ray Conference, Rio de
Janeiro, Brasil,  2--9 July   2013.

\bibitem[Aharonian \em{et~al.}(2006)Aharonian, Akhperjanian, Bazer-Bachi,
Beilicke, Benbow, Berge, Bernl{\"o}hr, Boisson, Bolz, Borrel,
et~al.]{aharonian2006observations}
Aharonian, F.; Akhperjanian, A.; Bazer-Bachi, A.; Beilicke, M.; Benbow, W.;
Berge, D.; Bernl{\"o}hr, K.; Boisson, C.; Bolz, O.; Borrel, V.;  et~al.
\newblock Observations of the Crab nebula with HESS.
\newblock {\em Astron. Astrophys.} {\bf 2006}, {\em 457},~899--915. [\href{http://dx.doi.org/10.1051/0004-6361:20065351}{CrossRef}]

\bibitem[Le~Bohec \em{et~al.}(1998)Le~Bohec, Degrange, Punch, Barrau,
Bazer-Bachi, Cabot, Chounet, Debiais, Dezalay, Djannati-Ata{\i}̈,
et~al.]{le1998new}
Le~Bohec, S.; Degrange, B.; Punch, M.; Barrau, A.; Bazer-Bachi, R.; Cabot, H.;
Chounet, L.; Debiais, G.; Dezalay, J.; Djannati-Ata\"{i}, A.;  et~al.
\newblock A new analysis method for very high definition Imaging Atmospheric
Cherenkov Telescopes as applied to the CAT telescope.
\newblock {\em Nucl. Instrum.   Methods Phys. Res. Sect. A  Accel. Spectrometers  Detect. Assoc. Equip.} {\bf 1998},
{\em 416},~425--437. [\href{http://dx.doi.org/10.1016/S0168-9002(98)00750-5}{CrossRef}]

\bibitem[De~Naurois and Rolland(2009)]{de2009high}
De~Naurois, M.; Rolland, L.
\newblock A high performance likelihood reconstruction of $\gamma$-rays for
imaging atmospheric Cherenkov telescopes.
\newblock {\em Astropart. Phys.} {\bf 2009}, {\em 32},~231--252. [\href{http://dx.doi.org/10.1016/j.astropartphys.2009.09.001}{CrossRef}]

\bibitem[Kertzman and Sembroski(1994)]{KERTZMAN1994629}
Kertzman, M.; Sembroski, G.
\newblock Computer simulation methods for investigating the detection
characteristics of TeV air Cherenkov telescopes.
\newblock {\em Nucl. Instrum.  Methods Phys. Res. Sect. A  Accel. Spectrometers  Detect. Assoc. Equip.} {\bf 1994},
{\em 343},~629--643. [\href{http://dx.doi.org/10.1016/0168-9002(94)90247-X}{CrossRef}]

\bibitem[Heck \em{et~al.}(1998)Heck, Knapp, Capdevielle, Schatz, Thouw,
et~al.]{heck1998corsika}
Heck, D.; Knapp, J.; Capdevielle, J.; Schatz, G.; Thouw, T.
\newblock CORSIKA: A Monte Carlo code to simulate extensive air {showers}.
\newblock {\em Rep. Fzka} {\bf 1998}, {\em 6019}. [\href{http://dx.doi.org/10.5445/IR/270043064}{CrossRef}]

\bibitem[Lemoine-Goumard \em{et~al.}(2006)Lemoine-Goumard, Degrange, and
Tluczykont]{lemoine2006selection}
Lemoine-Goumard, M.; Degrange, B.; Tluczykont, M.
\newblock Selection and 3D-reconstruction of gamma-ray-induced air showers with
a stereoscopic system of atmospheric Cherenkov telescopes.
\newblock {\em Astropart. Phys.} {\bf 2006}, {\em 25},~195--211. [\href{http://dx.doi.org/10.1016/j.astropartphys.2006.01.005}{CrossRef}]

\bibitem[Naumann-God{\'o} \em{et~al.}(2009)Naumann-God{\'o}, Lemoine-Goumard,
and Degrange]{naumann2009upgrading}
Naumann-God{\'o}, M.; Lemoine-Goumard, M.; Degrange, B.
\newblock Upgrading and testing the 3D reconstruction of gamma-ray air showers
as observed with an array of Imaging Atmospheric Cherenkov Telescopes.
\newblock {\em Astropart. Phys.} {\bf 2009}, {\em 31},~421--430. [\href{http://dx.doi.org/10.1016/j.astropartphys.2009.04.006}{CrossRef}]

\bibitem[Becherini \em{et~al.}(2011)Becherini, Djannati-Ata{\"\i}, Marandon,
Punch, and Pita]{becherini2011new}
Becherini, Y.; Djannati-Ata{\"\i}, A.; Marandon, V.; Punch, M.; Pita, S.
\newblock A new analysis strategy for detection of faint $\gamma$-ray sources
with Imaging Atmospheric Cherenkov Telescopes.
\newblock {\em Astropart. Phys.} {\bf 2011}, {\em 34},~858--870. [\href{http://dx.doi.org/10.1016/j.astropartphys.2011.03.005}{CrossRef}]

\bibitem[Parsons and Hinton(2014)]{parsons2014monte}
Parsons, R.; Hinton, J.
\newblock A Monte Carlo template based analysis for air-Cherenkov arrays.
\newblock {\em Astropart. Phys.} {\bf 2014}, {\em 56},~26--34. [\href{http://dx.doi.org/10.1016/j.astropartphys.2014.03.002}{CrossRef}]

\bibitem[Bernl{\"o}hr(2008)]{bernlohr2008simulation}
Bernl{\"o}hr, K.
\newblock Simulation of imaging atmospheric Cherenkov telescopes with CORSIKA
and sim\_telarray.
\newblock {\em Astropart. Phys.} {\bf 2008}, {\em 30},~149--158. [\href{http://dx.doi.org/10.1016/j.astropartphys.2008.07.009}{CrossRef}]

\bibitem[Vincent(2015)]{vincent2015monte}
Vincent, S.
\newblock A Monte Carlo template-based analysis for very high definition
imaging atmospheric Cherenkov telescopes as applied to the VERITAS telescope
array.
\newblock {\em arXiv} {\bf 2015}, arXiv:1509.01980.

\bibitem[Holler \em{et~al.}(2020)Holler, Lenain, de~Naurois, Rauth, and
Sanchez]{holler2020run}
Holler, M.; Lenain, J.P.; de~Naurois, M.; Rauth, R.; Sanchez, D.
\newblock A run-wise simulation and analysis framework for Imaging Atmospheric
Cherenkov Telescope arrays.
\newblock {\em Astropart. Phys.} {\bf 2020}, {\em 123},~102491. [\href{http://dx.doi.org/10.1016/j.astropartphys.2020.102491}{CrossRef}]

\bibitem[Bock \em{et~al.}(2004)Bock, Chilingarian, Gaug, Hakl, Hengstebeck,
Ji{\v{r}}ina, Klaschka, Kotr{\v{c}}, Savick{\`y}, Towers,
et~al.]{bock2004methods}
Bock, R.; Chilingarian, A.; Gaug, M.; Hakl, F.; Hengstebeck, T.; Ji{\v{r}}ina,
M.; Klaschka, J.; Kotr{\v{c}}, E.; Savick{\`y}, P.; Towers, S.;  et~al.
\newblock Methods for multidimensional event classification: A case study using
images from a Cherenkov gamma-ray telescope.
\newblock {\em Nucl. Instrum.  Methods Phys. Res. Sect. A Accel. Spectrometers Detect. Assoc. Equip.} {\bf 2004},
{\em 516},~511--528. [\href{http://dx.doi.org/10.1016/j.nima.2003.08.157}{CrossRef}]

\bibitem[Breiman \em{et~al.}(2017)Breiman, Friedman, Olshen, and
Stone]{breiman2017classification}
Breiman, L.; Friedman, J.H.; Olshen, R.A.; Stone, C.J.
\newblock {\em Classification and Regression Trees}; Routledge: Abingdon, UK,  2017.

\bibitem[Ohm \em{et~al.}(2009)Ohm, van Eldik, and Egberts]{ohm2009gamma}
Ohm, S.; van Eldik, C.; Egberts, K.
\newblock $\gamma$/hadron separation in very-high-energy $\gamma$-ray astronomy
using a multivariate analysis method.
\newblock {\em Astropart. Phys.} {\bf 2009}, {\em 31},~383--391. [\href{http://dx.doi.org/10.1016/j.astropartphys.2009.04.001}{CrossRef}]

\bibitem[Fiasson \em{et~al.}(2010)Fiasson, Dubois, Lamanna, Masbou, and
Rosier-Lees]{fiasson2010optimization}
Fiasson, A.; Dubois, F.; Lamanna, G.; Masbou, J.; Rosier-Lees, S.
\newblock Optimization of multivariate analysis for IACT stereoscopic systems.
\newblock {\em Astropart. Phys.} {\bf 2010}, {\em 34},~25--32. [\href{http://dx.doi.org/10.1016/j.astropartphys.2010.04.006}{CrossRef}]

\bibitem[Krause \em{et~al.}(2017)Krause, Pueschel, and
Maier]{krause2017improved}
Krause, M.; Pueschel, E.; Maier, G.
\newblock Improved $\gamma$/hadron separation for the detection of faint
$\gamma$-ray sources using boosted decision trees.
\newblock {\em Astropart. Phys.} {\bf 2017}, {\em 89},~1--9. [\href{http://dx.doi.org/10.1016/j.astropartphys.2017.01.004}{CrossRef}]

\bibitem[Albert \em{et~al.}(2008)Albert, Aliu, Anderhub, Antoranz, Armada,
Asensio, Baixeras, Barrio, Bartko, Bastieri,
et~al.]{albert2008implementation}
Albert, J.; Aliu, E.; Anderhub, H.; Antoranz, P.; Armada, A.; Asensio, M.;
Baixeras, C.; Barrio, J.; Bartko, H.; Bastieri, D.;  et~al.
\newblock Implementation of the random forest method for the imaging
atmospheric Cherenkov telescope MAGIC.
\newblock {\em Nucl. Instrum.  Methods Phys. Res. Sect. A Accel. Spectrometers Detect. Assoc. Equip.} {\bf 2008},
{\em 588},~424--432. [\href{http://dx.doi.org/10.1016/j.nima.2007.11.068}{CrossRef}]

\bibitem[Colin \em{et~al.}(2009)Colin, Tridon, Carmona, De~Sabata, Gaug,
Lombardi, Majumdar, Moralejo, Scalzotto, and Sitarek]{colin2009performance}
Colin, P.; Tridon, D.B.; Carmona, E.; De~Sabata, F.; Gaug, M.; Lombardi, S.;
Majumdar, P.; Moralejo, A.; Scalzotto, V.; Sitarek, J.
\newblock Performance of the MAGIC telescopes in stereoscopic mode.
\newblock {\em arXiv} {\bf 2009},    arXiv:0907.0960.

\bibitem[Goodfellow \em{et~al.}(2016)Goodfellow, Bengio, and
Courville]{goodfellow2016deep}
Goodfellow, I.; Bengio, Y.; Courville, A.
\newblock {\em Deep Learning}; MIT Press: Cambridge, MA, USA,  2016.

\bibitem[Feng \em{et~al.}(2016)Feng, Lin, Collaboration,
et~al.]{feng2016analysis}
Feng, Q.; Lin, T.T.; VERITAS Collaboration.
\newblock The analysis of VERITAS muon images using convolutional neural
networks.
\newblock {\em Proc. Int. Astron. Union} {\bf 2016},
{\em 12},~173--179. [\href{http://dx.doi.org/10.1017/S1743921316012734}{CrossRef}]

\bibitem[LeCun \em{et~al.}(2015)LeCun, Bengio, and Hinton]{lecun2015deep}
LeCun, Y.; Bengio, Y.; Hinton, G.
\newblock Deep learning.
\newblock {\em Nature} {\bf 2015}, {\em 521},~436--444. [\href{http://dx.doi.org/10.1038/nature14539}{CrossRef}] [\href{http://www.ncbi.nlm.nih.gov/pubmed/26017442}{PubMed}]

\bibitem[Shilon \em{et~al.}(2019)Shilon, Kraus, B{\"u}chele, Egberts, Fischer,
Holch, Lohse, Schwanke, Steppa, and Funk]{shilon2019application}
Shilon, I.; Kraus, M.; B{\"u}chele, M.; Egberts, K.; Fischer, T.; Holch, T.L.;
Lohse, T.; Schwanke, U.; Steppa, C.; Funk, S.
\newblock Application of deep learning methods to analysis of imaging
atmospheric Cherenkov telescopes data.
\newblock {\em Astropart. Phys.} {\bf 2019}, {\em 105},~44--53. [\href{http://dx.doi.org/10.1016/j.astropartphys.2018.10.003}{CrossRef}]

\bibitem[Mangano \em{et~al.}(2018)Mangano, Delgado, Bernardos, Lallena,
V{\'a}zquez, Consortium, et~al.]{mangano2018extracting}
Mangano, S.; Delgado, C.; Bernardos, M.I.; Lallena, M.; V{\'a}zquez, J.J.R.;
CTA Consortium.
\newblock Extracting gamma-ray information from images with convolutional
neural network methods on simulated cherenkov telescope array data.
\newblock In \emph{IAPR Workshop on Artificial Neural Networks in Pattern Recognition};
Springer: Cham, Switzerland,  2018; pp. 243--254.

\bibitem[Nieto~Castaño \em{et~al.}(2017)Nieto~Castaño, Brill, Kim, and
Humensky]{NietoCasta}
Nieto~Castaño, D.; Brill, A.; Kim, B.; Humensky, T.B.
\newblock {Exploring deep learning as an event classification method for the
Cherenkov Telescope {Array}}.
\newblock  In~{Proceedings of the} 35th International Cosmic Ray Conference,  (ICRC2017), Busan, Korea,  12--20 July 2017; p.~809. 

\bibitem[Holch \em{et~al.}(2017)Holch, Shilon, Büchele, Fischer, Funk,
Groeger, Jankowsky, Lohse, Schwanke, and Wagner]{Holch}
Holch, T.L.; Shilon, I.; Büchele, M.; Fischer, T.; Funk, S.; Groeger, N.;
Jankowsky, D.; Lohse, T.; Schwanke, U.; Wagner, P.
\newblock {Probing Convolutional Neural Networks for Event Reconstruction in
Gamma-Ray Astronomy with Cherenkov {Telescopes}}.
\newblock In~{Proceedings of the} 35th International Cosmic Ray Conference,  (ICRC2017), Busan, Korea,  12--20 July 2017; p. 795.

\bibitem[{Cash}(1979)]{1979ApJ...228..939C}
{Cash}, W.
\newblock {Parameter estimation in astronomy through application of the
likelihood ratio.}
\newblock {\em Astrophys. J.} {\bf 1979}, {\em 228},~939--947. [\href{http://dx.doi.org/10.1086/156922}{CrossRef}]

\bibitem[D'Agostini(1998)]{d1998jeffreys}
D'Agostini, G.
\newblock Jeffreys priors versus experienced physicist priors-arguments against
objective Bayesian theory.
\newblock {\em arXiv} {\bf 1998},  	arXiv:physics/9811045.

\bibitem[Aharonian \em{et~al.}(2001)Aharonian, Akhperjanian, Barrio,
Bernl{\"o}hr, B{\"o}rst, Bojahr, Bolz, Contreras, Cortina, Denninghoff,
et~al.]{aharonian2001evidence}
Aharonian, F.; Akhperjanian, A.; Barrio, J.; Bernl{\"o}hr, K.; B{\"o}rst, H.;
Bojahr, H.; Bolz, O.; Contreras, J.; Cortina, J.; Denninghoff, S.;  et~al.
\newblock Evidence for TeV gamma ray emission from Cassiopeia A.
\newblock {\em Astron. Astrophys.} {\bf 2001}, {\em 370},~112--120. [\href{http://dx.doi.org/10.1051/0004-6361:20010243}{CrossRef}]

\bibitem[Rowell(2003)]{rowell2003new}
Rowell, G.P.
\newblock A new template background estimate for source searching in TeV
$\gamma$-ray astronomy.
\newblock {\em Astron. Astrophys.} {\bf 2003}, {\em 410},~389--396. [\href{http://dx.doi.org/10.1051/0004-6361:20031194}{CrossRef}]

\bibitem[Fernandes \em{et~al.}(2014)Fernandes, Horns, Kosack, Raue, and
Rowell]{fernandes2014new}
Fernandes, M.V.; Horns, D.; Kosack, K.; Raue, M.; Rowell, G.
\newblock A new method of reconstructing VHE $\gamma$-ray spectra: The Template
Background Spectrum.
\newblock {\em Astron. Astrophys.} {\bf 2014}, {\em 568},~A117. [\href{http://dx.doi.org/10.1051/0004-6361/201323156}{CrossRef}]

\bibitem[Berge \em{et~al.}(2007)Berge, Funk, and Hinton]{berge2007background}
Berge, D.; Funk, S.; Hinton, J.
\newblock Background modelling in very-high-energy $\gamma$-ray astronomy.
\newblock {\em Astron. Astrophys.} {\bf 2007}, {\em 466},~1219--1229. [\href{http://dx.doi.org/10.1051/0004-6361:20066674}{CrossRef}]

\bibitem[Rolke and Lopez(2001)]{rolke2001confidence}
Rolke, W.A.; Lopez, A.M.
\newblock Confidence intervals and upper bounds for small signals in the
presence of background noise.
\newblock {\em Nucl. Instrum.  Methods Phys. Res. Sect. A  Accel. Spectrometers Detect. Assoc. Equip.} {\bf 2001},
{\em 458},~745--758. [\href{http://dx.doi.org/10.1016/S0168-9002(00)00935-9}{CrossRef}]

\bibitem[{Li} and {Ma}(1983)]{Li_Ma}
{Li}, T.P.; {Ma}, Y.Q.
\newblock {Analysis methods for results in gamma-ray astronomy.}
\newblock {\em Astrophys. J.} {\bf 1983}, {\em 272},~317--324.\linebreak161295. [\href{http://dx.doi.org/10.1086/161295}{CrossRef}]

\bibitem[Cousins \em{et~al.}(2008)Cousins, Linnemann, and
Tucker]{cousins2008evaluation}
Cousins, R.D.; Linnemann, J.T.; Tucker, J.
\newblock Evaluation of three methods for calculating statistical significance
when incorporating a systematic uncertainty into a test of the
background-only hypothesis for a Poisson process.
\newblock {\em Nucl. Instrum.  Methods Phys. Res. Sect. A  Accel. Spectrometers Detect. Assoc. Equip.} {\bf 2008},
{\em 595},~480--501. [\href{http://dx.doi.org/10.1016/j.nima.2008.07.086}{CrossRef}]

\bibitem[Linnemann(2003)]{linnemann2003measures}
Linnemann, J.T.
\newblock Measures of significance in HEP and astrophysics.
\newblock {\em arXiv} {\bf 2003},    arXiv:physics/0312059.

\bibitem[Vianello(2018)]{vianello2018significance}
Vianello, G.
\newblock The significance of an excess in a counting experiment: Assessing the
impact of systematic uncertainties and the case with a Gaussian background.
\newblock {\em  Astrophys. J. Suppl. Ser.} {\bf 2018}, {\em
236},~17. [\href{http://dx.doi.org/10.3847/1538-4365/aab780}{CrossRef}]

\bibitem[Klepser(2012)]{klepser2012generalized}
Klepser, S.
\newblock A generalized likelihood ratio test statistic for Cherenkov telescope
data.
\newblock {\em Astropart. Phys.} {\bf 2012}, {\em 36},~64--76. [\href{http://dx.doi.org/10.1016/j.astropartphys.2012.04.008}{CrossRef}]

\bibitem[Nievas-Rosillo and Contreras(2016)]{nievas2016extending}
Nievas-Rosillo, M.; Contreras, J.
\newblock Extending the Li\&Ma method to include PSF information.
\newblock {\em Astropart. Phys.} {\bf 2016}, {\em 74},~51--57.

\bibitem[Klepser(2017)]{klepser2017optimal}
Klepser, S.
\newblock The optimal on-source region size for detections with counting-type
telescopes.
\newblock {\em Astropart. Phys.} {\bf 2017}, {\em 89},~10--13. [\href{http://dx.doi.org/10.1016/j.astropartphys.2017.01.005}{CrossRef}]

\bibitem[Weiner(2015)]{weiner2015new}
Weiner, O.M.
\newblock A new time-dependent likelihood technique for detection of gamma-ray
bursts with IACT arrays.
\newblock {\em arXiv} {\bf 2015}, arXiv:1509.01290.

\bibitem[Dickinson and Conrad(2013)]{dickinson2013handling}
Dickinson, H.; Conrad, J.
\newblock Handling systematic uncertainties and combined source analyses for
Atmospheric Cherenkov Telescopes.
\newblock {\em Astropart. Phys.} {\bf 2013}, {\em 41},~17--30. [\href{http://dx.doi.org/10.1016/j.astropartphys.2012.10.004}{CrossRef}]

\bibitem[Spengler(2015)]{spengler2015significance}
Spengler, G.
\newblock Significance in gamma ray astronomy with systematic errors.
\newblock {\em Astropart. Phys.} {\bf 2015}, {\em 67},~70--74. [\href{http://dx.doi.org/10.1016/j.astropartphys.2015.02.002}{CrossRef}]

\bibitem[{Loredo}(1992)]{Loredo}
{Loredo}, T.J.
\newblock {Promise of Bayesian inference for astrophysics.}
\newblock  In \emph{Statistical Challenges in Modern Astronomy}; Springer: New York, NY, USA, 1992; pp. 275--297.\_31. [\href{http://dx.doi.org/10.1007/978-1-4613-9290-3_31}{CrossRef}]

\bibitem[D’Amico \em{et~al.}(2021)D’Amico, Terzi{\'c},
Stri{\v{s}}kovi{\'c}, Doro, Strzys, and van Scherpenberg]{d2021signal}
D’Amico, G.; Terzi{\'c}, T.; Stri{\v{s}}kovi{\'c}, J.; Doro, M.; Strzys, M.;
van Scherpenberg, J.
\newblock Signal estimation in on/off measurements including event-by-event
variables.
\newblock {\em Phys. Rev. D} {\bf 2021}, {\em 103},~123001. [\href{http://dx.doi.org/10.1103/PhysRevD.103.123001}{CrossRef}]

\bibitem[Gregory(2005)]{gregory2005bayesian}
Gregory, P.
\newblock {\em Bayesian Logical Data Analysis for the Physical Sciences: A
Comparative Approach with Mathematica\textsuperscript{\textregistered} Support}; Cambridge
University Press: Cambridge, UK,  2005.

\bibitem[Knoetig(2014)]{knoetig2014signal}
Knoetig, M.L.
\newblock Signal discovery, limits, and uncertainties with sparse on/off
measurements: An objective bayesian analysis.
\newblock {\em  Astrophys. J.} {\bf 2014}, {\em 790},~106. [\href{http://dx.doi.org/10.1088/0004-637X/790/2/106}{CrossRef}]

\bibitem[Jeffreys(1998)]{jeffreys1998theory}
Jeffreys, H.
\newblock {\em The Theory of Probability}; OUP: Oxford, UK,  1998.

\bibitem[Casadei(2014)]{casadei2014objective}
Casadei, D.
\newblock Objective Bayesian analysis of “on/off” measurements.
\newblock {\em  Astrophys. J.} {\bf 2014}, {\em 798},~5. [\href{http://dx.doi.org/10.1088/0004-637X/798/1/5}{CrossRef}]

\bibitem[Rolke \em{et~al.}(2005)Rolke, Lopez, and Conrad]{rolke2005limits}
Rolke, W.A.; Lopez, A.M.; Conrad, J.
\newblock Limits and confidence intervals in the presence of nuisance
parameters.
\newblock {\em Nucl. Instrum.  Methods Phys. Res. Sect. A  Accel. Spectrometers Detect. Assoc. Equip.} {\bf 2005},
{\em 551},~493--503. [\href{http://dx.doi.org/10.1016/j.nima.2005.05.068}{CrossRef}]

\bibitem[O'Hagan and Forster(2004)]{o2004kendall}
O'Hagan, A.; Forster, J.J.
\newblock {\em Kendall's Advanced Theory of Statistics, Volume 2B: Bayesian
Inference}; Arnold: London, UK,  2004; Volume~2, 

\bibitem[Ackermann \em{et~al.}(2012)Ackermann, Ajello, Albert, Allafort,
Atwood, Axelsson, Baldini, Ballet, Barbiellini, Bastieri,
et~al.]{ackermann2012fermi}
Ackermann, M.; Ajello, M.; Albert, A.; Allafort, A.; Atwood, W.; Axelsson, M.;
Baldini, L.; Ballet, J.; Barbiellini, G.; Bastieri, D.;  et~al.
\newblock The Fermi large area telescope on orbit: Event classification,
instrument response functions, and calibration.
\newblock {\em  Astrophys. J. Suppl. Ser.} {\bf 2012}, {\em
203},~4. [\href{http://dx.doi.org/10.1088/0067-0049/203/1/4}{CrossRef}]

\bibitem[Albert \em{et~al.}(2007)Albert, Aliu, Anderhub, Antoranz, Armada,
Asensio, Baixeras, Barrio, Bartko, Bastieri, et~al.]{albert2007unfolding}
Albert, J.; Aliu, E.; Anderhub, H.; Antoranz, P.; Armada, A.; Asensio, M.;
Baixeras, C.; Barrio, J.; Bartko, H.; Bastieri, D.;  et~al.
\newblock Unfolding of differential energy spectra in the MAGIC experiment.
\newblock {\em Nucl. Instrum.  Methods Phys. Res. Sect. A Accel. Spectrometers Detect. Assoc. Equip.} {\bf 2007},
{\em 583},~494--506. [\href{http://dx.doi.org/10.1016/j.nima.2007.09.048}{CrossRef}]

\bibitem[Tikhonov(1963)]{tikhonov1963solution}
Tikhonov, A.N.
\newblock \emph{On the Solution of Ill-Posed Problems and the Method of
Regularization};
\newblock  Doklady Akademii Nauk; Russian Academy of Sciences: Moscow, Russia,  1963; Volume 151,
pp. 501--504.

\bibitem[Bertero(1989)]{bertero1989linear}
Bertero, M.
\newblock Linear inverse and III-posed problems.
\newblock {\em Adv. Electron. Electron Phys.} {\bf 1989}, {\em
75},~1--120.

\bibitem[Schmelling(1994)]{schmelling1994method}
Schmelling, M.
\newblock The method of reduced cross-entropy A general approach to unfold
probability distributions.
\newblock {\em Nucl. Instrum.  Methods Phys. Res. Sect. A Accel. Spectrometers Detect. Assoc. Equip.} {\bf 1994},
{\em 340},~400--412. [\href{http://dx.doi.org/10.1016/0168-9002(94)90119-8}{CrossRef}]

\bibitem[D'Agostini(1995)]{d1995multidimensional}
D'Agostini, G.
\newblock A multidimensional unfolding method based on Bayes' theorem.
\newblock {\em Nucl. Instrum.  Methods Phys. Res. Sect. A Accel. Spectrometers Detect. Assoc. Equip.} {\bf 1995},
{\em 362},~487--498. [\href{http://dx.doi.org/10.1016/0168-9002(95)00274-X}{CrossRef}]

\bibitem[D'Agostini(2010)]{d2010improved}
D'Agostini, G.
\newblock Improved iterative Bayesian unfolding.
\newblock {\em arXiv} {\bf 2010},    arXiv:1010.0632.

\bibitem[Choudalakis(2012)]{choudalakis2012fully}
Choudalakis, G.
\newblock Fully bayesian unfolding.
\newblock {\em arXiv} {\bf 2012},    arXiv:1201.4612.

\bibitem[Rico(2020)]{rico2020gamma}
Rico, J.
\newblock Gamma-Ray Dark Matter Searches in Milky Way Satellites---A
Comparative Review of Data Analysis Methods and Current Results.
\newblock {\em Galaxies} {\bf 2020}, {\em 8},~25. [\href{http://dx.doi.org/10.3390/galaxies8010025}{CrossRef}]

\bibitem[Acciari \em{et~al.}(2020)Acciari, Ansoldi, Antonelli, Engels, Baack,
Babi{\'c}, Banerjee, de~Almeida, Barrio, Gonz{\'a}lez,
et~al.]{acciari2020bounds}
Acciari, V.A.; Ansoldi, S.; Antonelli, L.A.; Engels, A.A.; Baack, D.;
Babi{\'c}, A.; Banerjee, B.; de~Almeida, U.B.; Barrio, J.A.; Gonz{\'a}lez,
J.B.;  et~al.
\newblock Bounds on Lorentz invariance violation from MAGIC observation of GRB
190114C.
\newblock {\em Phys. Rev. Lett.} {\bf 2020}, {\em 125},~021301. [\href{http://dx.doi.org/10.1103/PhysRevLett.125.021301}{CrossRef}] [\href{http://www.ncbi.nlm.nih.gov/pubmed/32701326}{PubMed}]

\bibitem[Weekes \em{et~al.}(1989)Weekes, Cawley, Fegan, Gibbs, Hillas, Kowk,
Lamb, Lewis, Macomb, Porter, et~al.]{weekes1989observation}
Weekes, T.C.; Cawley, M.F.; Fegan, D.; Gibbs, K.; Hillas, A.; Kowk, P.; Lamb,
R.; Lewis, D.; Macomb, D.; Porter, N.;  et~al.
\newblock Observation of TeV gamma-rays from the crab nebula using the
atmospheric cherenkov imaging technique.
\newblock {\em Astrophys. J.} {\bf 1989}, {\em 342},~379--395. [\href{http://dx.doi.org/10.1086/167599}{CrossRef}]

\bibitem[Amato and Olmi(2021)]{amato2021crab}
Amato, E.; Olmi, B.
\newblock The Crab Pulsar and Nebula as seen in gamma-rays.
\newblock {\em Universe} {\bf 2021}, {\em 7},~448. [\href{http://dx.doi.org/10.3390/universe7110448}{CrossRef}]

\bibitem[doi(2019)]{doi:10.1142/10986}
Al Samarai, I.; Batista, R.A.; de Almeida, U.B.; dos Anjos, R.D.C.; Dal Pino, E.D.G.; de los Reyes Lopez, R.; de Ona Wilhelmi, E.; Dominis Prester, D.; Freixas Coromina, L.; Garcia López, R.; et al. {Science with the Cherenkov Telescope Array}. \emph{Astrophys. J. Suppl.}  \textbf{2019}, \emph{240}, 32. [\href{http://dx.doi.org/10.1142/10986}{CrossRef}]

\end{thebibliography}
\end{document}